\documentclass[11pt]{article}
\usepackage{jheppub}
\usepackage{bm}
\usepackage{mathtools}

\usepackage{tikz}
\usetikzlibrary{intersections, calc,positioning,decorations.pathreplacing,decorations.pathmorphing}

\def\CB{{\cal B}}
\def\CC{{\cal C}}
\def\CD{{\cal D}}

\def\CM{{\cal M}}
\def\CN{{\cal N}}
\def\CO{{\cal O}}

\def\BR{\mathbb{R}}
\def\BS{\mathbb{S}}

\def\BG{\mathbf{G}}
\def\BP{\mathbf{P}}
\def\BQ{\mathbf{Q}}
\def\SO{\mathrm{SO}}
\def\GL{\mathrm{GL}}
\def\FD{\mathfrak{D}}
\def\Fd{\mathfrak{d}}
\def\sH{\mathsf{H}}

\def\sQ{\mathsf{Q}}

\def\sV{\mathsf{V}}
\def\wc{\circ}
\def\bc{\bullet}

\title{Spinning conformal defects}

\author[a,b]{Nozomu Kobayashi}
\author[a]{and Tatsuma Nishioka}

\affiliation[a]{Department of Physics, Faculty of Science,
University of Tokyo,\\
Bunkyo-ku, Tokyo 113-0033, Japan}
\affiliation[b]{Kavli Institute for the Physics and Mathematics of the Universe (WPI), \\
The University of Tokyo Institutes for Advanced Study, The University of Tokyo, \\
Kashiwa, Chiba 277-8583, Japan}

\abstract{
We consider conformal defects with spins under the rotation group acting on the transverse directions.
They are described in the embedding space formalism in a similar manner to spinning local operators, and their correlation functions with bulk and defect local operators are determined by the conformal symmetry.
The operator product expansion (OPE) structure of spinning conformal defects is examined by decomposing it into the spinning defect OPE block that packages all the contribution from a conformal multiplet.
The integral representation of the block derived in the shadow formalism is facilitated to deduce recursion relations for correlation functions of two spinning conformal defects.
In simple cases, we construct spinning defect correlators by acting differential operators recursively on scalar defect correlators.
}

\preprint{UT-18-11, IPMU18-0084}

\begin{document}
\maketitle

\section{Introduction}
Quantum field theories (QFTs) allow not only for local operators but also extended objects like line and surface operators defined on one- and two-dimensional manifolds respectively \cite{Gukov:2008sn,Gukov:2014gja}.
The latter class of operators of general (co)dimensions are called defects and play distinguished roles in probing non-local phenomena that are inaccessible by local operators in QFTs.
For example, a $q$-form symmetry is generated by an extended operator associated with a codimension-$(q+1)$ manifold $\CM^{(d-q-1)}$ in $d$ dimensions.
The charged objects are also extended on $q$-dimensional manifolds $\CC^{(q)}$ surrounded by $\CM^{(d-q-1)}$, then both the generators and charged operators are defects for $q>0$.
Higher-form symmetries allow a refined classification of phases in QFTs under the spontaneous symmetry breaking and place non-trivial constraints on renormalization group flows \cite{Gaiotto:2014kfa}.
Another situation where defects come into play in an unexpected way is entanglement entropy in QFT that is commonly defined for a spacial region at a given time slice.
It has an alternative description as a codimension-two defect bounding the region \cite{Calabrese:2004eu,Cardy:2013nua,Hung:2014npa,Bianchi:2015liz}, which can be made more manifest in a supersymmetric analogue of entanglement R{\'e}nyi entropies \cite{Nishioka:2013haa,Nishioka:2016guu}.

While a complete classification of defects is far from our reach so far, there are a particularly nice class of defects called conformal defects that preserve a part of the conformal symmetry in conformal field theories (CFTs).
The existence of boundaries and defects brings about new structures in the operator product expansion (OPE), which was examined originally in boundary CFTs \cite{Cardy:1984bb,McAvity:1995zd} and have attracted renewed attention in general defect CFTs more recently \cite{Billo:2016cpy,Gadde:2016fbj,Fukuda:2017cup,Lauria:2017wav,Armas:2017pvj} along with the developments of the bootstrap programs \cite{Liendo:2012hy,Gliozzi:2015qsa,Gaiotto:2013nva,Liendo:2016ymz}.
There are also a certain type of anomalies associated with boundaries and defects that have begun to be classified in the recent studies
 \cite{Jensen:2015swa,Herzog:2015ioa,Fursaev:2015wpa,Solodukhin:2015eca,Jensen:2017eof,Herzog:2017kkj,Herzog:2017xha}.

A conformal defect of codimension-$m$ is either a flat hyperplane $\BR^{d-m}$ or a hypersphere $\BS^{d-m}$ and breaks the $d$-dimensional conformal group $\SO(d+1,1)$ to the subgroup $\SO(d-m+1,1) \times \SO(m)$ acting on the worldvolume of the defect and the transverse directions as the $(d-m)$-dimensional conformal group and the transverse rotation, respectively.
Defect CFTs accommodate defect local operators $\hat \CO$ in addition to bulk local operators $\CO$.
Defect local operators are characterized by primary operators in a $(d-m)$-dimensional CFT supported on a defect with extra transverse spins under the $\SO(m)$ group.
Similarly, conformal defects of codimension-$m$ themselves are labeled by the transverse spins under the $\SO(m)$ group as they can be constructed by smearing a defect local operator with a transverse spin over the worldvolumes.
Correlation functions involving bulk and defect local primary operators are constrained by the residual symmetry as demonstrated by \cite{Gadde:2016fbj,Billo:2016cpy} for scalar conformal defects.
It is worthwhile to introduce the transverse spins into conformal defects and examine the consequences for their correlation functions.

In this paper, we consider a conformal defect with the transverse spin-$s$ and study their correlation functions with bulk and defect local operators.\footnote{Conformal defects in the other representations may be constructed in a similar manner to the case for local operators \cite{Costa:2014rya,Costa:2016hju,Rejon-Barrera:2015bpa}.}
Our implementation is based on the embedding space formalism \cite{Dirac:1936fq,Weinberg:2010fx,Costa:2011mg} that allows a systematic construction of symmetric traceless tensors in CFT.
The formalism has been adapted to describing scalar conformal defects and defect primary operators in \cite{Gadde:2016fbj,Billo:2016cpy}, and will be extended to incorporating the defect spins in the present paper.
We further explore the OPE of a spinning conformal defect in terms of bulk local operators by introducing the spinning defect OPE (DOPE) block that is a projection of the defect onto a conformal multiplet as in the case for a scalar defect \cite{Fukuda:2017cup}.
The shadow formalism \cite{Ferrara:1973eg,Ferrara:1972xe,Ferrara:1973vz,Ferrara:1972uq,Ferrara:1972ay,SimmonsDuffin:2012uy} is exploited to derive an integral representation of the spinning DOPE block, which turns out to be useful to deduce differential equations for the blocks from those for the one-point functions of bulk local operators in the presence of a spinning defect.
Interestingly, we find recursion relations associating 
the spin-$s$ DOPE blocks to the spin-$(s-2)$ blocks acted by differential operators, which bears a resemblance to the recursion relations for three-point functions of bulk local operators \cite{Costa:2011dw}.
These are inherited to recursion relations for the correlators of two spinning conformal defects through the integral representation of the spinning DOPE blocks, and we find it possible to reduce the spinning defect correlators to scalar defect correlators in certain cases.
We believe an abstract approach adopted by \cite{Karateev:2017jgd} would provide a more systematic derivation for the recursion relations between spinning defect correlators.

This paper is organized as follow.
In section \ref{ss:embedding} we quickly review the embedding space formalism for spinning local operators and defect local operators.
Section \ref{ss:spinning_defect} describes the formalism adapted to scalar defects, followed by an extension to spinning conformal defects.
In section \ref{ss:correlators} we determine a few class of correlation functions of bulk and defect local operators in the presence of a spinning defect.
Unlike the scalar case, the bulk one-point function is not necessarily fixed uniquely due to the defect spin.
For conserved currents, the conservation law provides additional constraints between the OPE coefficients.
We show that a spinning defect of codimension-$d$ may be viewed as a special type of a pair of spinning local operators by comparing their correlators explicitly.
We then construct differential operators raising the defect spin by two and derive recursion relations between the one-point functions with different defect spins.
Section \ref{ss:SDOPE} examines the OPE structure of spinning conformal defects that is efficiently encoded into the spinning DOPE blocks we shall introduce.
We deduce the integral representation of the spinning DOPE blocks in the shadow formalism, and employ it to translate the recursion relations for the bulk one-point functions into those for the blocks.
Section \ref{ss:2-pt_of_defects} is concerned with the two-point functions of spinning defects. 
We show they are calculable by solving the quadratic Casimir equation, but we alternatively develop the recursive method to construct the spinning defect correlator from scalar defect correlators.
Finally we discuss the implications of our results and conclude with future directions for this program in section \ref{ss:discussion}. 

\medskip

\noindent
{\bf Notes added:} After submitting this manuscript to arXiv, we became aware of \cite{Guha:2018snh} which has an overlap with our work.

\section{Embedding space formalism} \label{ss:embedding}
Throughout this paper, we consider a Euclidean $d$-dimensional CFT in flat space $\BR^d$.
The conformal group $\SO(d+1,1)$ acts on this space non-linearly, but it can be linearized as the Lorentz transformation in embedding space $\BR^{d+1,1}$ with the coordinates $X^M~(M=1,\cdots, d+2)$ and the $\SO (d+1,1)$ invariant inner product,
\begin{align}
	X\cdot Y \equiv \BG_{MN}\,X^M\, Y^N = \sum_{i=1}^{d+1}X^i\, Y^i - X^{d+2}\, Y^{d+2} \ .
\end{align}
A CFT is thought of as living on a $d$-dimensional subspace called the projective null cone,
\begin{align}
	X^2 \equiv X\cdot X= 0\ ,\qquad X \sim \lambda\, X\qquad (\lambda \in \BR)\ ,
\end{align}
which is manifestly $\SO(d+1,1)$ invariant.
The physical coordinates $x^\mu~(\mu = 1,\cdots d)$ of $\BR^d$ can be lifted to the embedding coordinates $X^M$ in various ways.
One realization is the Poincar\'e section that is nicely written in the light cone coordinates $X^+ = X^{d+1} + X^{d+2}, X^- = X^{d+1} - X^{d+2}$ as,
\begin{align}
	X^M = (X^+,X^-,X^\mu) = (1,x^2,x^\mu) \ .
\end{align}
Then the distance squared between two points $x_1$ and $x_2$ in the physical space is given by the inner product of the two embedding vectors $X_1$, $X_2$, 
\begin{align}
	X_{12}\equiv -2 X_1\cdot X_2 = (x_1 - x_2)^2 \ .
\end{align}

\subsection{Spinning bulk local operators} 
Let us apply the above formalism to describing primary operators.
We restrict our attention to symmetric traceless spin-$l$ primary fields with dimension $\Delta$.
In order to lift up a tensor field $\phi_{\mu_1 \cdots \mu_l}(x)$ into the embedding space we consider a tensor field $\Phi_{M_1 \cdots M_l}(X)$ satisfying the following conditions \cite{Costa:2011mg}:
\begin{enumerate}
    \item living on light cone $X^2=0$, 
    \item \label{traceless} symmetric and traceless tensor,
    \item \label{pure_gauge} defined modulo tensors such that $X_{M_i} F_{M_1 \cdots \hat{M}_i \cdots M_l}$,\footnote{$F_{M_1 \cdots \hat{M}_i \cdots M_l} $ is an arbitrary tensor with the index $M_i$ eliminated. These terms will vanish when projected onto the Poincar\'e section.}
    \item transverse, $ X^M \Phi_{M M_2 \cdots M_l} = 0$,
    \item homogeneous of degree $-\Delta$: $\Phi_{M_1 \cdots M_l}(\lambda \,X) = \lambda^{-\Delta}\,\Phi_{M_1 \cdots M_l}(X)$.
\end{enumerate}
The original tensor field $\phi_{\mu_1 \cdots \mu_l}(x)$ can be recovered by pulling back the lifted tensor $\Phi_{M_1 \cdots M_l}(X)$ on the Poincar\'e section,
\begin{align}
    \phi_{\mu_1 \cdots \mu_l}(x) = \left.\frac{\partial X^{M_1}}{\partial x^{\mu_1}} \cdots \frac{\partial X^{M_l}}{\partial x^{\mu_l}}\, \Phi_{M_1 \cdots M_l}(X) \right|_{X^M = (1,x^2,x^i)} \ .
\end{align}

These conditions are most economically written in encoding spinning local operators into polynomials by contracting their indices with an auxiliary vector $Z$,
\begin{align}
	\CO_{\Delta, l} (X, Z) \equiv \CO_{\Delta, M_1\cdots M_l} (X)\, Z^{M_1}\cdots Z^{M_l} \ .
\end{align}
To ensure $\CO_{\Delta, M_1\cdots M_l} (X)$ being  symmetric traceless transverse tensor, the auxiliary vector must be transverse to the coordinate vector $X$ and itself,
\begin{align}
	Z\cdot X = Z^2 = 0 \ .
\end{align}
These conditions are invariant under the ``gauge" symmetry $Z\to Z + \lambda\, X$ for any constant $\lambda$.
It will be convenient to introduce the gauge invariant antisymmetric tensor $C_{ZX}$ by
\begin{align}
	(C_{ZX})^{MN} \equiv Z^M X^N - X^M Z^N \ .
\end{align}
Then correlation functions can depend on $Z$ only through the gauge invariant tensor. 
The gauge invariance is equivalently represented as the transversality condition,
\begin{align}
	X\cdot \partial_Z\,\CO_{\Delta, l} (X, Z) = 0 \ ,
\end{align}
and the dimension and spin are written as the homogeneity conditions,
\begin{align}\label{Homogeneity}
	X\cdot \partial_X\, \CO_{\Delta, l} (X, Z) = -\Delta\,\CO_{\Delta, l} (X, Z)\ , \qquad Z\cdot \partial_Z\, \CO_{\Delta, l} (X, Z) = l\,\CO_{\Delta, l} (X, Z) \ .
\end{align}

In this formulation, the conformal generators $\hat J_{MN}$ are realized as a differential operator acting on the encoding polynomial $\CO_{\Delta, l}(X,Z)$,
\begin{align}
	\hat J_{MN} = X_M\,\frac{\partial}{\partial X^N} - X_N\,\frac{\partial}{\partial X^M} + Z_M\,\frac{\partial}{\partial Z^N} - Z_N\,\frac{\partial}{\partial Z^M} \ .
\end{align}
It follows that $\CO_{\Delta, l}$ is an eigenstate of the quadratic Casimir operator $\hat J^2 \equiv \hat J_{MN}\hat J^{MN}/2$,\footnote{In our convention, the conformal generators are anti-hermitian.}
\begin{align}
	\begin{aligned}
	 \hat J^2\,\CO_{\Delta, l} (X, Z) = - \CC_{\Delta, l}\,\CO_{\Delta, l} (X, Z) \ , \qquad \CC_{\Delta, l} = \Delta (\Delta - d) + l(l+d -2) \ .
\end{aligned}
\end{align}

The original spinning primary $\CO_{M_1 \cdots M_l}(X)$ can be recovered  by acting the Todorov differential operator $D_M$ on the encoding polynomial $\CO_l (X,Z)$  \cite{Dobrev:1975ru}:
\begin{align}
    D_M &= \left(\frac{d-2}{2} + Z \cdot \frac{\partial}{\partial Z} \right)\frac{\partial}{\partial Z^M}-\frac{1}{2}Z_M \frac{\partial^2}{\partial Z \cdot \partial Z} \ ,\\
    \CO_{M_1 \cdots M_l}(X)  &= \frac{1}{l!(d/2 -1)_l}D_{M_1} \cdots D_{M_l}\,\CO_l(X,Z) \ ,
\end{align}
where $(a)_l = \Gamma(a+l)/\Gamma(a)$ is the Pochhammer symbol.
This is useful when we compute the contraction of two tensors $f^{\mu_1\cdots \mu_l}$ and $g^{\mu_1\cdots \mu_l}$ in embedding space.
The result is given by the product of their encoding polynomials $F(X,Z)$ and $G(X,Z)$ with replacing an auxiliary vector $Z$ with the Todorov operator $D$ for $F(X,Z)$ \cite{Costa:2011mg},
\begin{align}\label{Contracted_Tensors}
	f_{\mu_1\cdots \mu_l}(x)\, g^{\mu_1\cdots \mu_l} (x) = \frac{1}{l! (d/2 -1)_l}\, F(X, D)\, G(X, Z) \ .
\end{align}

\subsection{Spinning defect local operators}
A codimension-$m$ defect in CFT$_d$ preserves the $\SO(m) \times \SO(d-m+1,1)$ subgroup of the whole conformal symmetry $\SO(d+1,1)$.
The transverse rotation group $\SO(m)$ can be regarded as a flavor symmetry for the CFT$_{d-m}$ living on the defect, and the defect local operators $\hat \CO_{j,s}$ are labeled by two types of spins, the transverse spin $s$ for the $\SO(m)$ group and the parallel spin $j$ for the $\SO(d-m) \subset \SO (d-m+1,1)$, which are most simply described by the embedding space formalism \cite{Billo:2016cpy} as we will review soon.

The physical coordinates $x^\mu$ have parallel $x^a$ and transverse $x^i$ components to the defect,
\begin{align}
	x^\mu = (x^a, x^i) \ , \qquad (a = 1, \cdots, d-m, \quad i = 1, \cdots, m) \ .
\end{align}
To describe a defect local operator $\hat \CO_{j,s}(x^a)$ of parallel spin $j$ and transverse spin $s$, we introduce a polynomial of two auxiliary vectors $z^a$ and $w^i$,
\begin{align}
	\hat \CO_{j,s}(x^a, z^a, w^i) \equiv \hat \CO_{a_1\cdots a_j, i_1\cdots i_s}(x^a) \, z^{a_1} \cdots z^{a_j}\,w^{i_1} \cdots w^{i_s} \ .
\end{align}
For the coefficient $\CO_{a_1\cdots a_j, i_1\cdots i_s}(x^a)$ being a symmetric traceless tensor for both transverse and parallel coordinates, we impose the auxiliary vectors to be null,
\begin{align}
	z^a z_a = 0 \ , \qquad w^i w_i = 0 \ .
\end{align}

Now we consider the uplift of defect local operators to the embedding space.
The physical coordinates on the defect with $x^i = 0$ are lifted to a null vector $\hat X^M$ with null parallel components $X^A$,
\begin{align}
	\hat X^M = (X^A, x^i = 0) \ , \qquad X^A X_A = 0 \ , \qquad (A = 1, \cdots d-m + 2) \ .
\end{align}
Namely, $X^A$ is the embedding space vector for CFT$_{d-m}$ on the defect.
In the Poincar{\'e} section, the embedding vector is represented by
\begin{align}
	\hat X^M = (1, x^a x_a, x^a, x^i = 0) \ .
\end{align}
The embedding vector $\hat Z$ for the parallel auxiliary vector $z^a$ can be fixed by the transversality conditions,
\begin{align}
	\hat X\cdot \hat Z = 0 \ , \qquad \hat Z^2 = 0 \ .
\end{align}
Solving them in the Poincar{\'e} section yields
\begin{align}
	\hat Z^M = (0, 2x^a z_a , z^a , z^i = 0 ) \ , \qquad z^a z_a = 0 \ .
\end{align}
Finally we uplift the transverse auxiliary vector $w^i$ to
\begin{align}
	\hat W^M = (0,0, w^a = 0, w^i ) \ , \qquad w^i w_i = 0 \ .
\end{align}
It is a null vector orthogonal to the defect,
\begin{align}\label{Transverse_Vector}
	\hat X\cdot \hat W = 0 \ , \qquad \hat W^2 = 0 \ ,\qquad \hat Z \cdot \hat W = 0 \ . 
\end{align}
With these preparations, a defect local operator with spins is encoded into the polynomial in the index-free notation,
\begin{align}
	\hat \CO_{j,s} (\hat X, \hat Z, \hat W) \equiv \hat \CO_{M_1\cdots M_j, N_1\cdots N_s} (\hat X)\, \hat Z^{M_1} \cdots \hat Z^{M_j}\, \hat W^{N_1}\cdots \hat W^{N_s} \ .
\end{align}
As in the case of the bulk local operators, correlation functions involving defect local operators are subject to the gauge invariance under the shift $\hat Z \rightarrow \hat Z + \lambda\, \hat X$.
Hence $\hat Z$ always appears in the gauge invariant form,
\begin{align}
\hat C^{MN}  \equiv C^{MN}_{\hat Z \hat X} = \hat Z^M \hat X^N - \hat X^M \hat Z^N \ .
\end{align}

\section{Spinning conformal defects} \label{ss:spinning_defect}
The embedding space formalism is adapted to a scalar defect in two different ways  by \cite{Billo:2016cpy} and \cite{Gadde:2016fbj}.
After reviewing the two approaches to calculate  correlation functions in a scalar defect CFT in section \ref{ss:defects_in_embedding}, we  will comment on their relation in section \ref{ss:correlators_in_scalar_defect}.
We will then extend these approaches to implement spinning conformal defects in section \ref{ss:spinning_defect_embedding}.

\subsection{Defects in embedding space} \label{ss:defects_in_embedding}
Conformal defects can be described in the embedding space formalism in two equivalent ways, one as a hypersurface in the projective null cone specified by the normal vectors and the other as a worldvolume theory with extra flavor indices.

\begin{figure}
\centering
\begin{tikzpicture}[thick,scale=1.4]
			\draw[very thick, orange, fill=orange!20] (0,0) --++ (1,1.5) --++ (0,-3) --++ (-1,-1.5) --cycle;
            \draw[->] (0.5, -1) --++ (1.5,0) node[above] {$P_\alpha$};
            \node at (0.5, 2) {\large $\CD^{(m)}(P_\alpha)$};
            \node at (0.5, -1.8) {$X^A$};
            \coordinate (A) at (4.5,0) node at (A) [below] {\large $\CO(Y_1)$};
            \fill (A) circle [radius=1pt];
            \coordinate (B) at (0.5,-0.2) node at (B) [below] {\large $\hat \CO(\hat Y_2)$};
            \fill (B) circle [radius=1pt];
		\end{tikzpicture}
        \caption{A conformal defect is a hypersurface orthogonal to the frame vectors $P_\alpha$. Defect local operators $\hat \CO$ are supported on the worldvolume while bulk local operators $\CO$ can be placed anywhere.}
        \label{fig:defect_embedding}
\end{figure}
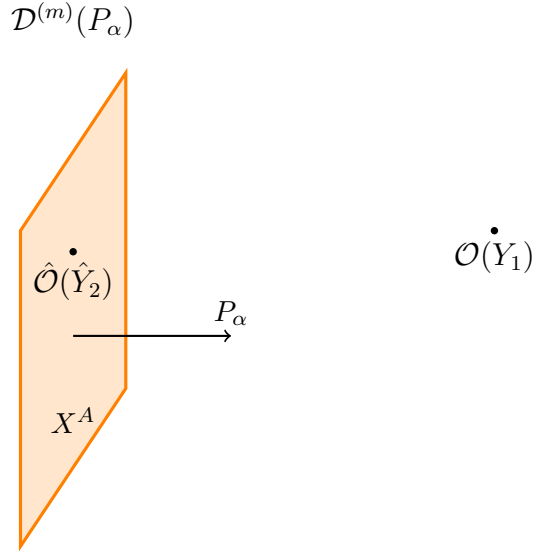

First we begin with the approach by Gadde \cite{Gadde:2016fbj} who fixes the location $X$ of a codimension-$m$ defect $\CD^{(m)}$ by specifying a set of the unit normal vectors $P_\alpha~(\alpha = 1,\cdots,m)$ in the projective null cone (see figure \ref{fig:defect_embedding}),
\begin{align}
	P_\alpha \cdot X = 0 \ ,\qquad X^2 = 0 \ , \qquad P_\alpha \cdot P_\beta = \delta_{\alpha\beta} \ .
\end{align}
The correlators of the defect are determined by imposing the residual conformal invariance and the $\GL(m)$ invariance for the frame vectors $P_\alpha$.
In the frame vector approach, the conformal transformation acting on defects is realized as differential operators,
\begin{align}\label{Orbital_generator}
	L_{MN}(P_\alpha) =  P_{\alpha, M} \,\frac{\partial}{\partial P_\alpha^N} - P_{ \alpha, N} \,\frac{\partial}{\partial P_\alpha^M} \ .
\end{align}

The other equivalent approach introduced by Bill\'o et.\,al \cite{Billo:2016cpy} splits the embedding coordinates $X^M$ into the parallel and transverse directions,
\begin{align}
    M = (A,I)\ , \qquad A = 1,\,2,\,\cdots,\,d-m+2\ ,\quad I = 1,\cdots,m\ ,
\end{align}
where a flat codimension-$m$ defect is described as a hypersurface located at $X^I=0$.
The conformal group $\SO(d+1,1)$ is manifestly broken to the subgroups $\SO(d-m+1,1)$ and $\SO(m)$, each of which acts on the parallel and transverse planes respectively.
In this index-split approach, 
there are two types of scalar products for two embedding space vectors $X^M$, $Y^M$,
\begin{align}\label{Splitting_inner_product}
    X \bc Y \equiv  \bm{\eta}_{AB}\,X^A \,Y^B\ , \qquad X \wc Y \equiv \bm{\delta}_{IJ}\,X^I \,Y^J\ ,
\end{align}
contracted with the induced metrics $\bm{\eta}_{AB} \equiv \text{diag}(+,+,\cdots,+,-)$ and $\bm{\delta}_{IJ} \equiv \text{diag}(+,+,\cdots,+)$ on and normal to the defect.
When the inner product of two vectors vanishes, $X\cdot Y = 0$, 
these two inner products are not independent and can be interchanged with each other,
\begin{align}
    X \bc Y = - X \wc Y \ .
\end{align}

One can easily translate the frame vector $P_\alpha$ into the inner products $\wc$ and $\bc$ through the $\SO(m)$ invariant matrix,
\begin{align}\label{TransverseProjectors}
	\BP^{MN} \equiv \sum_\alpha P_\alpha^M\,P_\alpha^N \ .
\end{align}
This is the projection operator onto the transverse plane to the defect and
one finds the dictionary between the inner products \eqref{Splitting_inner_product} and the projection operator,
\begin{align}\label{Gadde_to_Billo}
\begin{aligned}
	\BP_{MN}\,X^M\, Y^N & \quad & \longleftrightarrow  & &\quad  X \wc Y \ , \\
    (\BG -\BP)_{MN}\, X^M\, Y^N & \quad & \longleftrightarrow & &\quad  X \bc Y\ .
\end{aligned}
\end{align}
We will mostly use the index-split approach in calculating the correlators of spinning operators in the presence of defects, while we will switch to the frame vector description to make manifest the action of the residual conformal group on the correlators.

\subsection{Correlators in scalar defects} \label{ss:correlators_in_scalar_defect}
In a defect CFT, the correlation functions of bulk local operators should be calculated in the presence of a defect,
\begin{align}
 \langle \CO(X_1) \cdots \CO(X_k) \rangle_{\CD} = 
 \langle \CD^{(m)}\, \CO(X_1) \cdots \CO(X_k)  \rangle \ .
\end{align}
These types of correlators can be fixed  by the conformal symmetry and homogeneity in the embedding space formalism in parallel with the correlators of local operators.

\paragraph{Frame vector approach}
We use the frame vectors $P_\alpha$ to impose the conformal symmetry on the correlators.
To illustrate how it works in a simplest example, let us consider the correlation function of a scalar defect and a spin-$l$ operator,
\begin{align}
	\langle \CD^{(m)}(P_\alpha)\, \CO_{\Delta, l}(X, Z)\rangle \ .
\end{align}
This correlator should be a scalar function with correct dimensions in the index-free notation.
The scalar invariants respecting the gauge redundancy that we can construct out of the vectors $P_\alpha, X$ and $Z$ are $P_\alpha \cdot X$ and $P_\alpha\cdot C_{ZX}\cdot P_\beta$.
In addition, we must contract the frame index $\alpha$ to make them invariant under the $\SO(m)$ symmetry.
Hence the following two invariants are allowed to show up in the correlator,
\begin{align}
	(P^\alpha \cdot X)(P_\alpha \cdot X) \ , \qquad (P^\alpha \cdot C_{ZX} \cdot P^\beta) (P_\alpha \cdot C_{ZX} \cdot P_\beta) \ .
\end{align}
By taking into account the homogeneity \eqref{Homogeneity} of the spin-$l$ operator, the correlator must be a homogeneous function with 
\begin{itemize}
    \item degree $-\Delta$ in $X$,
    \item degree $l$ in $Z$.
\end{itemize}
These conditions fix the form of the correlator uniquely up to a factor \cite{Billo:2016cpy,Costa:2011mg,SimmonsDuffin:2012uy}:
\begin{align}\label{One_Point_Defect}
\begin{aligned}
	\langle \CD^{(m)}(P_\alpha)\, \CO_{\Delta, l}(X, Z)\rangle &= \frac{a_{\Delta, l}}{\left[ (P^\alpha \cdot X)(P_\alpha \cdot X)\right]^{(\Delta + l)/2}} \left[ (P^\beta \cdot C_{ZX} \cdot P^\gamma)(P_\beta \cdot C_{ZX} \cdot P_\gamma)\right]^{l/2}\ ,\\
    &= \frac{a_{\Delta, l}}{\left[ (P^\alpha \cdot X)(P_\alpha \cdot X)\right]^{(\Delta + l)/2}} \left[ (Z \cdot C_{P^\beta P^\gamma} \cdot X)(Z \cdot C_{P_\beta P_\gamma} \cdot X)\right]^{l/2}\ ,
\end{aligned}
\end{align}
where we used the relation $P^\beta \cdot C_{ZX} \cdot P^\gamma = Z \cdot C_{P^\beta P^\gamma} \cdot X$ in going from the first line to the second.
The correlators with non-zero spin vanish for $m=1$ as there is only one frame vector $P_1$ resulting in $C_{P_1 P_1} = 0$.
This is consistent with the result in boundary CFT (see e.g.\,\cite{Liendo:2012hy}).
Note that the correlators are parity invariant and make sense only for even $l$ \cite{Billo:2016cpy}. 
In other words there are no non-vanishing parity invariant correlators for odd $l$.\footnote{Parity odd correlators can be built with the $\SO(d+1,1)$-invariant $\epsilon$-tensor \cite{Costa:2011mg,Billo:2016cpy,Fukuda:2017cup}.}

\paragraph{Index split approach}
In this approach, the gauge invariant tensor $C^{MN} \equiv C_{ZX}^{MN}$ is decomposed into three tensors $C^{AB},C^{AI}$, and $C^{IJ}$ in the presence of defects.
In order to construct gauge invariants, however, we only need the one with mixed indices,
\begin{align}
    C^{AI} = Z^A X^I -X^A Z^I\ .
\end{align}
This is because the other two, $C^{AB}$ and $C^{IJ}$, can be written as linear combinations of $C^{AI}$ \cite{Billo:2016cpy},
\begin{align}
    C^{AB}\, Q_A\, R_B &= \frac{X \bc R}{X \wc S}\,C^{AI}\, Q_A\, S_I - \frac{X \bc Q}{X \wc S}\,C^{AI}\,R_A\, S_I \ ,\\
    C^{IJ}\, Q_I\, R_J &= \frac{X \wc Q}{X \bc S}\,C^{AI}\, S_A\, R_I - \frac{X \wc R}{X \bc S}\,C^{AI}\, S_A\, Q_I\ ,   
\end{align}
for arbitrary vectors $Q,R,S$.\footnote{We can always take $S^M = X^M$ to simplify the expressions.} Furthermore the following identity holds for the gauge invariant tensor:
\begin{align}
    C^{AI}\,C_{BI}\,C^{BJ} = \frac{1}{2}(C^{BI}C_{BI})\,C^{AJ}\ .
\end{align}
Thus linearly-independent scalar invariants contain at most two $C^{AI}$s.

Let us calculate $\langle \CD^{(m)} \, \CO_{\Delta,l}(X,Z) \rangle$ in this approach.
The gauge invariant tensor $C^{AI}$ can be contracted with either $X$ or itself.
It is easy to see that $C^{AI}C_{AI}$ is only the linearly-independent invariant.
Imposing the homogeneity, the correlator is fixed to be
\begin{align}\label{One_Point_Defect_Billo}
    \langle \CD^{(m)} \, \CO_{\Delta,l}(X,Z) \rangle = \frac{(C^{AI}C_{AI})^{l/2}}{(X \wc X)^{(\Delta + l)/2}}\ .
\end{align}
In order to compare it with the correlator in the frame vector approach, we use the dictionary \eqref{Gadde_to_Billo} adapted to the present case,
\begin{align}
	X \wc X \quad \longleftrightarrow \quad \BP_{MN}\, X^M\, X^N = (P^\alpha\cdot X) (P_\alpha\cdot X) \ ,
\end{align}
and
\begin{align}\label{CC_to_PP}
	\begin{aligned}
		C^{AI}C_{AI} &= 2\left[ (Z\bc Z)(X \wc X) - (Z\bc X) (X \wc Z) \right] \\
        	&\longleftrightarrow\quad - \,(P^\alpha \cdot C_{ZX} \cdot P^\beta)(P_\alpha \cdot C_{ZX} \cdot P_\beta) \\
            &\qquad\qquad = -\, (Z\cdot C_{P^\alpha P^\beta}\cdot X)(Z\cdot C_{P_\alpha P_\beta}\cdot X)\ .
	\end{aligned}
\end{align}
Then it is clear \eqref{One_Point_Defect} and \eqref{One_Point_Defect_Billo} are equivalent up to a factor.

Throughout the rest of this paper, we will use the $\wc$ and $\bc$ notation for conformal invariants when applicable.
For instance, the gauge invariant contracted with two vectors $Q$ and $R$ is represented by
\begin{align}
	C^{AI}\,Q_A\,R_I = Q\bc C \wc R \ ,
\end{align}
in this notation.
Similarly we write the concatenated gauge invariants as
\begin{align}
	C^{AI}C_{AJ} = (C\bc C)^I_{~J}\ , \qquad C^{AI}C_{BI} = (C\wc C)^A_{~B}\ .
\end{align}

\subsection{Spinning defects in embedding space} \label{ss:spinning_defect_embedding}

\begin{figure}
\centering
       \begin{tikzpicture}
            \draw[very thick, orange] (0,1.5) node[above, black] {\large $\BR^{d-m}$} -- (0,-1.5) node[below, black] {\large $\CD^{(m)}_s$};
            \draw[->, thick] (-0.5,0.5) arc [start angle=120, end angle=420, x radius=1cm, y radius=0.5cm];
            \node at (-2,-0.8) {\large $\SO(m)$};
        \end{tikzpicture}
        \caption{A spinning defect of codimension-$m$ has a transverse spin under the rotation group $\SO(m)$ acting on the transverse directions.}
        \label{fig:spinnig_defect}
\end{figure}
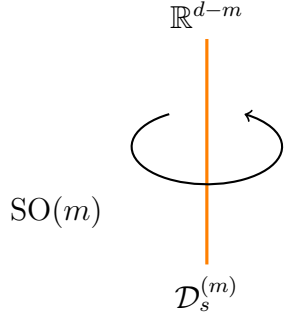

In what follows, we will adapt the index-free notation to spinning conformal defects in the embedding space (figure \ref{fig:spinnig_defect}).
We contract the spin indices of a spin-$s$ conformal defect $\CD^{(m)}_s$ with an auxiliary vector $\hat W$,
\begin{align}\label{Spinning_Defect}
	\CD^{(m)}_s(P_\alpha, \hat W) \equiv \CD^{(m)}_{M_1\cdots M_s}(P_\alpha)\, \hat W^{M_1} \cdots \hat W^{M_s} \ .
\end{align}
Then the spin indices are automatically symmetrized.
In order to implement spin indices for the transverse group $\SO(m)$, we must impose the transversality condition for the auxiliary null vector $\hat W$ to the defect,
\begin{align}\label{TransverseNullVector}
	\BP_{MN}\, \hat W^M\, \hat W^N = 0 \ , \qquad \BP^{MN}\,\hat W_N = \hat W^M \ .
\end{align}
Then the orthogonality $X\cdot \hat W = 0$ follows from the second condition.\footnote{
The first condition is equivalently written as $(P_\alpha \cdot \hat W)^2 = 0$, hence it has a non-trivial solution, $P_\alpha\cdot \hat W \neq 0$, if $\hat W$ is complex in Euclidean signature \cite{Dobrev:1977qv}.
It can have a non-trivial real solution in Lorentzian signature as one of the frame vectors is timelike.
}
Note that these conditions correspond to \eqref{Transverse_Vector} for defect local operators.
They are invariant under the symmetry translating $\hat W$ along the frame vector $P_\alpha$,
\begin{align}
	\hat W \to \hat W + \lambda\, P_\alpha \ .
\end{align}
Hence correlation functions including a spinning defect $\CD^{(m)}_s$ enjoy the same symmetry.

For completeness let us write a spinning defect in the index-split approach,
\begin{align}
	\CD^{(m)}_s(\hat W) \equiv \CD^{(m)}_{I_1\cdots I_s}\, \hat W^{I_1} \cdots \hat W^{I_s} \ ,
\end{align}
where $\hat W$ is a transverse vector $\hat W_I$, satisfying
\begin{align}
	\hat W \wc \hat W = 0 \ .
\end{align}
The orthogonality condition may be written as $X\wc \hat W = 0$.

\section{Correlation functions with a spinning defect} \label{ss:correlators}
We have formulated a spinning conformal defect in embedding space in the previous section.
Here we will turn to the constraints imposed by the conformal symmetry on their correlation functions with bulk and defect local operators.
In section \ref{ss:Recursion_one-point}, we will introduce differential operators increasing defect spins by two when acted on spinning defects, and present recursion relations between the one-point functions of bulk spinning operators with different defect spins.

\subsection{Bulk one-point function}
Let us consider the one-point function of a bulk spin-$l$ operator $\CO_{\Delta,l}$ in the presence of a spin-$s$ defect,
\begin{align}
    \langle \CD^{(m)}_s(\hat W)\, \CO_{\Delta,l}(X,Z) \rangle \ .
\end{align}
This is a homogeneous function of degree $-\Delta$, $l$ and $s$ in $X$, $Z$ and $\hat W$, respectively.

In order to fix the correlator, 
we need to find an independent basis of scalar invariants built from the embedding vectors $\hat W^I, X^M$ and $Z^M$.
In the index-split approach, we can construct the $\wc$ and $\bc$ products, but the $\wc$ product can be used for $\hat W$.
They are also related by $X\bc X = - X\wc X$ as $X$ is a null vector.
Hence without $Z$, there are two independent invariants,
\begin{align}
	X \wc X  \ , \qquad \sQ_1 = \frac{X\wc \hat W}{(X \wc X)^{1/2}} \ .
\end{align}
On the other hand, invariants including $Z$ should be built out of the anti-symmetric tensor $C^{AI}$, then there are two independent gauge invariants,
\begin{align}
    \sQ_2 = \frac{X \bc C \wc \hat{W}}{X \wc X}\ , \qquad \sQ_3 = \frac{C^{AI}\,C_{AI}}{X \wc X} \ .
\end{align}
There are more gauge invariants using two gauge invariant tensors $C^{AI}$ with one of the indices contracted, but they are not independent of $\sQ_2$ because of the identities,
\begin{align}\label{C_identities}
	\begin{aligned}
		C^{AI}\, C_{A}^{~J}&= - \frac{C^{AI}\,C^{BJ}\,X_A\,X_B}{(X\wc X)} + \frac{(C^{AI}\,C_{AI})X^I\,X^J}{2(X\wc X)} \ ,\\
		C^{AI}\,C^B_{~~I} &= \frac{C^{AI}C^{BJ}X_I X_J}{(X \wc X)} - \frac{(C^{AI}\,C_{AI})X^A\,X^B}{2(X\wc X)}  \ .
    \end{aligned}
\end{align}
Thus the correlator is a polynomial of the four invariants, $X\wc X,\, Q_1,\, Q_2,\, Q_3$, satisfying the homogeneity condition
\begin{align}\label{bulk_one_point}
    \langle \CD^{(m)}_s(\hat W)\, \CO_{\Delta,l}(X,Z) \rangle = \sum_n\, N_n\,\frac{ \sQ_1^{s-l + 2n}\, \sQ_2^{l-2n}\, \sQ_3^{n}}{(X \wc X)^{\Delta /2}} \ ,
\end{align}
where $N_n$ are undetermined constants and the range of the non-negative integer $n$ is determined by the non-negativity of the exponents of $\sQ_i$s,
\begin{align}\label{Range_of_n}
	 \text{max}\left( 0, \left\lceil \frac{l-s}{2} \right\rceil \right) \le n \le \left\lfloor\frac{l}{2}\right\rfloor \ .
\end{align}
As a non-trivial check, the correlator reproduces the previous result \eqref{One_Point_Defect_Billo} for $s=0$.

For a codimension-$1$ defect, the invariant $\sQ_3$ vanishes as seen from the relation \eqref{CC_to_PP} where $C_{P_\alpha P_\beta}$ becomes zero due to the antisymmetric property and $P_\alpha = P_1$.
Moreover, it must vanish unless $s=0$ as there are no spins for a trivial group $\SO(m=1)$.
It may also be accounted for by the condition $(P_1\cdot \hat W)^2=0$ that forces the auxiliary vector $\hat W$ to identically zero even in Lorentzian signature, resulting in no invariants other than $X\wc X$.
Hence the non-vanishing one-point function for a codimension-one defect is 
\begin{align}\label{bulk_one_point_codim_one}
    \langle \CD^{(1)}\, \CO_{\Delta}(X) \rangle = \frac{1}{(X \wc X)^{\Delta /2}} \ ,
\end{align}
where we use the unit normalization.

\subsubsection{Codimension-$d$ defects}
A conformal defect of codimension-$d$ is special as it consists of a pair of points.
Let us denote the locations of the pair by $X_1$ and $X_2$, then they can be used to span the dual frame vectors $\tilde{P}_{\tilde\alpha}$ $(\tilde\alpha = 1,2)$ for a codimension-$d$ defect,
\begin{align}
	\tilde P_1 = \frac{X_1 - X_2}{X_{12}} \ , \qquad \tilde P_2 = \frac{X_1 + X_2}{X_{12}} \ ,
\end{align}
satisfying the orthogonality condition $\tilde P_{\tilde\alpha} \cdot \tilde P_{\tilde\beta} = \eta_{\tilde\alpha \tilde\beta}$ for $\eta_{\tilde\alpha \tilde\beta} = \text{diag} (1, -1)$. 
The $\SO(1,1)$ symmetry acting on the dual frame vectors is equivalent to the invariance under the simultaneous scaling of $X_1$ and $X_2$,
\begin{align}\label{ScaleInv}
	X_1 \to \lambda\, X_1 \ , \qquad X_2 \to \lambda^{-1}\, X_2 \ .
\end{align}

The projection operator $\tilde\BP$ onto the worldvolume of the defect can be written in terms of the position vectors $X_i~(i=1,2)$ as,
\begin{align}
\tilde\BP^{MN} = \sum_{\tilde \alpha=1,2}\tilde{P}^{\tilde \alpha\, M} \tilde{P}_{\tilde\alpha}^N = -2\,\frac{X_1^M X_2^N + X_2^M X_1^N}{X_{12}} \ .
\end{align}
We can use them to translate the basic invariants $\sQ_i$ in the correlator \eqref{bulk_one_point} into invariants constructed out of the vectors $X_1$ and $X_2$ through the dictionary,
\begin{align}\label{Dual_dictionary}
\begin{aligned}
	(\BG -\tilde\BP)_{MN}\,X^M\, Y^N & \quad & \longleftrightarrow  & &\quad  X \wc Y \ , \\
    \tilde\BP_{MN}\, X^M\, Y^N & \quad & \longleftrightarrow & &\quad  X \bc Y\ .
\end{aligned}
\end{align}
In this case, the orthogonality condition for $\hat{W}$ amounts to
\begin{align}\label{orthogonal_condition_for_codim-d}
X_i \cdot \hat{W} = 0 \qquad (i=1,2)\ ,\qquad \tilde \BP \cdot \hat W = 0 \ .
\end{align}

We are now in position to represent the correlator \eqref{bulk_one_point} in terms of the position vectors $X_i$.
Given the relations \eqref{Dual_dictionary} and \eqref{orthogonal_condition_for_codim-d},
the invariant $\sQ_1$ can be rewritten into
\begin{align}
	\begin{aligned}
	\sQ_1 &= \frac{X  \wc \hat{W}}{(X \wc X)^{1/2}}\ ,\\ 
    &= \frac{ X \cdot \hat{W}- X \cdot \tilde\BP \cdot \hat{W}}{(-X \cdot \tilde\BP \cdot X)^{1/2}} \ ,\\
	&= \frac{(X \cdot \hat{W})\, X_{12}^{1/2}}{2 (X\cdot X_1)^{1/2}(X \cdot X_2)^{1/2}}\ ,
	\end{aligned}
\end{align}
and similarly for $\sQ_2$ and $\sQ_3$,
\begin{align}
	\sQ_2 & = \frac{X \bc C \wc \hat{W}}{X \wc X} = - \frac{(X \cdot X_1)( X_2 \cdot C \cdot \hat{W}) + (X \cdot X_2)( X_1 \cdot C \cdot \hat{W})}{2 (X \cdot X_1)( X \cdot X_2)}\ , \\
	\sQ_3 & = \frac{C^{AI}C_{AI}}{X \wc X} = 2\,\frac{(X_1 \cdot C \cdot X_2)^2}{(X \cdot X_1) (X \cdot X_2) X_{12}}\ .
\end{align}
Interestingly, we find that $\sQ_3$ is built from a more elementary invariant $\tilde \sQ_3 \equiv \sQ_3^{1/2}$ when defects are of codimension-$d$.
We find it more convenient to introduce a new invariant,
\begin{align}
		\tilde \sQ_2 &\equiv - \frac{\sQ_1^2\,\sQ_3}{2} + \sQ_2^2 = \frac{(X_1 \cdot C \cdot \hat{W}) (X_2 \cdot C \cdot \hat{W})}{(X \cdot X_1)(X \cdot X_2)}\ ,
\end{align}
and expand the one-point function in the basis $(\sQ_1, \tilde \sQ_2, \tilde \sQ_3)$,
\begin{align}\label{onepoint_for_codim-d}
\langle \CD^{{(d)}}_s (X_1,X_2,\hat{W})\, \CO_{\Delta,l}(X,Z) \rangle = \sum_n N_n'\, \frac{\sQ_1^{s-l+n} \tilde \sQ_2^{(l-n)/2} \tilde \sQ_3^{n}}{(X_1 \cdot X_2)^{-\Delta/2} (X\cdot X_1)^{\Delta/2}( X\cdot X_2)^{\Delta/2}}\ .
\end{align}

\subsubsection{Codimension-$d$ defects and a pair of local operators} 
We shall regard the bulk one-point function \eqref{onepoint_for_codim-d} for a codimension-$d$ defect as a three-point function of a pair of local operators located at $X_1, X_2$ and an external operator $\CO_{\Delta,l} (X,Z)$.
Namely we want to associate a spinning defect of codimension-$d$ to a bilocal operator with spin under $\SO(d)$,\footnote{To be precise we should write 
\begin{align}\label{Defect_to_PairLocalOp}
	\begin{aligned}
		\CD^{(d)}_{a_1 \cdots a_s} &= \left[\Phi(X_1) \Phi (X_2) \right]_{a_1 \cdots a_s} \ ,\\
		& = \sum_{k=0}^{s} \Phi_{\{a_1 \cdots a_k }(X_1)\, \Phi_{ a_{k+1} \cdots a_s \} } (X_2) \ .
     \end{aligned}
\end{align}
} 
\begin{align}
	\begin{aligned}
 \CD^{(d)}_s(P_{\alpha},\hat{W}) &= \left[\Phi \Phi \right]_s (X_1,X_2,\hat{W})\ , \\
  &= \sum_{s_1,s_2} \Phi_{s_1}(X_1,\hat{W})\,\Phi_{s_2}(X_2,\hat{W})\ , \qquad (s_1 + s_2 = s) \ ,
	\end{aligned}
\end{align}
where the dimensions of $\Phi$'s must be zero so as to be consistent with the dimensionality of the defect.

In order for the correspondence between a spinning defect of codimension-$d$ and a pair of spinning local operators to work, the correlation function \eqref{onepoint_for_codim-d} has to be reproduced from the three-point function of spinning operators \cite{Costa:2011mg},
\begin{align}\label{General_form_of_3pt}
\langle \Phi_{s_1}(X_1,\hat W)\,\Phi_{s_2}(X_2,\hat W)\, \CO_{\Delta,l}(X,Z) \rangle = \frac{F(X_1,X_2,X,\hat W,Z)}{X_{12}^{(s-\Delta-l)/2}(X \cdot X_1)^{(\Delta + l +s_1 -s_2)/2}(X_2 \cdot X)^{(\Delta+l+s_2-s_1)/2}}\ .
\end{align}
The function $F$ is a polynomial of degree $s$ and $l$ in $\hat W$ and $Z$, respectively.
Let us enumerate the basic invariants for the polynomial $F$ that are constructed from the gauge invariant tensors $C_{X_i\hat W}$, $C_{XZ}$ \cite{Costa:2011mg},
\begin{align}
 \sH_i &= C_{X_i \hat W}\cdot C_{X Z}   = 2\, X_i \cdot C_{X Z} \cdot \hat W\ , \quad (i = 1,2) \ ,  \\
 \sV_1 &= -2\, \frac{X_2 \cdot C_{X_1 \hat W} \cdot X}{X\cdot X_2} = \frac{(X\cdot \hat W)\, X_{12}}{X \cdot X_2}\ ,  \\
 \sV_2 & = -2\, \frac{X_1 \cdot C_{X_2 \hat W} \cdot X}{X\cdot X_2}  = \frac{(X \cdot \hat W)\, X_{12}}{X \cdot X_1}\ , \\
 \sV & = \frac{X_1 \cdot C_{X Z} \cdot X_2}{X_{12}}\ . 
\end{align}
There are no invariants including $C_{X_1 \hat W} \cdot C_{X_2 \hat W}$ as they vanish due to the transverse conditions \eqref{orthogonal_condition_for_codim-d}.
These invariants are not linearly independent, and subject to the identity,
\begin{align}
	 \frac{\sH_1}{X\cdot X_1} - \frac{\sH_2}{X\cdot X_2} + 2 \frac{\sV_1\, \sV}{(X\cdot X_1)} = 0 \ .
\end{align}
Thus we can use $(\sH_1 \sH_2, \sV_1, \sV_2, \sV)$ as a linearly independent basis to expand $F$,
\begin{align}
 F(X_1,X_2,X,W,Z) = \sum\, a_{p\, m_1 m_2\, q}\, (\sH_1 \sH_2)^{p}\,\sV_1^{m_1}\,\sV_2^{m_2}\,\sV^{q}\ .
\end{align}

To match with the defect description in \eqref{onepoint_for_codim-d}, 
the correlator \eqref{General_form_of_3pt} should respect the $\SO(1,1)$ symmetry acting on $X_1$ and $X_2$ as a simultaneous rescaling \eqref{ScaleInv}, which yields the following constraint between the parameters,
\begin{align}
 m_1 - m_2 = s_1 - s_2 \ .
\end{align}
Furthermore, the homogeneity conditions for $\hat W$ and $Z$ give rise to two more constraint equations,
\begin{align}
2p + m_1 + m_2 &= s\ , \\
2p + q & = l\ .
\end{align}
Solving these constraints, the correlator can be written in the form,
\begin{align}
	&\sum a_{q}\, \frac{ (\sH_1 \sH_2)^{(l-q)/2}\,\sV_1^{(q-l)/2 + s_1}\,\sV_2^{(q-l)/2 + s_2}\,\sV^{q}}{X_{12}^{(s-\Delta-l)/2}(X \cdot X_1)^{(\Delta + l +s_1 -s_2)/2}(X_2 \cdot X)^{(\Delta+l+s_2-s_1)/2}}\ .
\end{align}
It appears to depend on the spins $s_1$ and $s_2$, but is actually independent of them.
Thus the summation over $s_1, s_2$ satisfying $s_1 + s_2 = s$ becomes
\begin{align}
	\begin{aligned}
	 \sum_{s_1 + s_2 = s} &\langle \Phi_{s_1}(X_1, \hat W)\,\Phi_{s_2}(X_2,\hat W)\,\CO_{\Delta,l}(X,Z) \rangle \\
     	& \qquad = \sum_q\, (s+1)\, 2^{l-q}\,a_{q}\, \frac{ \left[(X_1 \cdot C \cdot \hat{W})(X_2 \cdot C \cdot \hat{W}) \right]^{(l- q)/2}(X \cdot \hat{W})^{q-l+s}(X_1 \cdot C \cdot X_2)^{q}}{X_{12}^{(l-s-\Delta)/2}\left[(X \cdot X_1)(X \cdot X_2) \right]^{(\Delta + q + s )/2 }} \ .
     \end{aligned}
\end{align}
This is exactly the same form as the one-point function \eqref{onepoint_for_codim-d} with the identification $q=n$, which justifies the correspondence \eqref{Defect_to_PairLocalOp} between a codimension-$d$ defect and a pair of spinning local operators.

\subsection{Two-point function of defect local operators}
Now we consider the two-point function of defect local operators in the presence of a spinning defect,
\begin{align}
    \langle \CD^{(m)}_s (\hat W)\, \hat \CO_{\hat \Delta_1, j_1, s_1}(\hat X_1, \hat Z_1, \hat W_1) \, \hat \CO_{\hat \Delta_2, j_2, s_2}(\hat X_2, \hat Z_2, \hat W_2)\rangle \ .
\end{align}
We need an independent invariant basis for the correlator built out of the null vectors $\hat W^I, \hat X_i^A, \hat Z_i^A$ and $\hat W_i^I$ for $i=1,2$.
Recalling that $\hat X_i$ and $\hat W_i$ have non-zero components only in the parallel and transverse directions to the defect, respectively, one finds invariants without $\hat Z_i$,
\begin{align}
	\hat X_1 \bc \hat X_2 \ , \qquad \hat W_1 \wc \hat W_2 \ ,
\end{align}
for any $s$, and two more for $s\ge 1$,
\begin{align}
	\hat W \wc \hat W_i \quad (i = 1,2) \ .
\end{align}
Gauge invariants including $\hat Z_i$ must be built from $\hat C^{AB}_i$ in a similar manner to the two-point function of local operators in $(d-m)$-dimensional CFT.
There is only one non-vanishing invariant,
\begin{align}
    \hat C_1^{AB} \, \hat C_{2\,AB}\ .
\end{align}
The correlator is a homogeneous polynomial of degree $-\hat \Delta_i$, $j_i$, $s_i$ and $s$ in $X_i$, $\hat Z_i$, $\hat W_i$ and $\hat W$ for $i=1,2$, respectively.
This fixes the form uniquely, up to a factor,
\begin{align}
    \begin{aligned}
        &\langle \CD^{(m)}_s (\hat W)\, \hat \CO_{\hat \Delta_1, j_1, s_1}(\hat X_1, \hat Z_1, \hat W_1) \, \hat \CO_{\hat \Delta_2, j_2, s_2}(\hat X_2, \hat Z_2, \hat W_2)\rangle \\
            & \qquad \qquad = \delta_{\hat \Delta_1 \hat \Delta_2}\, \delta_{j_1 j_2}\, \frac{\left(\hat C_1^{AB} \, \hat C_{2,AB}\right)^{j_1}}{(\hat X_1\bc \hat X_2)^{\hat\Delta_1 + j_1}}  (\hat W\wc \hat W_1)^{(s_1 - s_2 + s)/2} (\hat W\wc \hat W_2)^{(s_2 - s_1 + s)/2} (\hat W_1\wc \hat W_2)^{(s_1 + s_2 - s)/2}\ .
    \end{aligned}
\end{align}
For a scalar defect with $s=0$, there are no invariants containing the auxiliary vector $\hat W$, hence the correlator becomes
\begin{align}
	\langle \CD^{(m)}\, \hat \CO_{\hat \Delta_1, j_1, s_1}(\hat X_1, \hat Z_1, \hat W_1) \, \hat \CO_{\hat \Delta_2, j_2, s_2}(\hat X_2, \hat Z_2, \hat W_2)\rangle 
       = \delta_{\hat \Delta_1 \hat \Delta_2}\, \delta_{j_1 j_2}\, \frac{\left(\hat C_1^{AB} \, \hat C_{2,AB}\right)^{j_1}}{(\hat X_1\bc \hat X_2)^{\hat\Delta_1 + j_1}}\, (\hat W_1\wc \hat W_2)^{(s_1 + s_2)/2}\ .
\end{align}

\subsection{Bulk-defect two-point function}
In defect CFT, the two-point function of a bulk and a defect local operators does not vanish in general.
Let us consider the two-point function in the presence of a spinning defect,
\begin{align}
    \langle \CD^{(m)}_s (\hat W)\, \CO_{\Delta_1, l_1}(X_1, Z_1) \, \hat \CO_{\hat \Delta_2, j_2, s_2}(\hat X_2, \hat Z_2, \hat W_2)\rangle \ .
\end{align}
There are two independent invariants without the auxiliary vectors,
\begin{align}
	X_1 \wc X_1 \ , \qquad X_1 \bc \hat X_2 \ .
\end{align}
Gauge invariants containing $\hat Z_2$ are built from only $\hat C_2^{AB}$, which can be contracted with $X_1$ or $Z_1$ through the other gauge invariant $C_1^{AB}$,
\begin{align}
    \sQ_{\mathsf{BD},1} = \frac{C_1^{AB}\, \hat C_{2,AB}}{(X_1\bc \hat X_2)}\ .
\end{align}
There are no other options for invariants with $C_2^{AB}$.

Moving to invariants without $\hat C_2$, one can contract $C_1^{AI}$ with itself,
\begin{align}
	\sQ_{\mathsf{BD},2} = \frac{C_1^{AI}\, C_{1,AI}}{(X_1\wc X_1)} \ .
\end{align}
There are several possibilities to contract $C_1^{AI}$ with $X_1,\hat X_2,\hat W_2 ,\hat W$.
The set of a linearly independent basis is given by\footnote{
For example, the invariants
\begin{align}
	\sQ_{\mathsf{BD},9} = \frac{\hat X_{2}\bc C_1\wc \hat W}{(X_1\bc \hat X_2)} \ , \qquad 
    \sQ_{\mathsf{BD},10} = \frac{\hat X_{2}\bc C_1\wc\hat W_{2}}{(X_1\bc \hat X_2)} \ ,
\end{align}
are not independent in this basis,
\begin{align}
	\sQ_{\mathsf{BD},9} = \sQ_{\mathsf{BD},3}\, \sQ_{\mathsf{BD},6} - \sQ_{\mathsf{BD},4} \ ,\qquad
	\sQ_{\mathsf{BD},10} = \sQ_{\mathsf{BD},3}\, \sQ_{\mathsf{BD},7} - \sQ_{\mathsf{BD},5} \ .
\end{align}
}
\begin{align}
    \sQ_{\mathsf{BD},3} &= \frac{\hat X_2 \bc C_1  \wc X_1}{(X_1\wc X_1)^{1/2}\,(X_1\bc \hat X_2)}\ , & \quad  
    \sQ_{\mathsf{BD},4} &= \frac{X_{1}\bc C_1 \wc \hat W}{(X_1\wc X_1)} \ , &\quad 
    \sQ_{\mathsf{BD},5} &= \frac{X_1 \bc C_1 \wc \hat W_2}{(X_1\wc X_1)}\ , \\
    \sQ_{\mathsf{BD},6} &= \frac{X_1 \wc \hat W}{(X_1\wc X_1)^{1/2}}\ ,& \quad 
    \sQ_{\mathsf{BD},7} &= \frac{X_1 \wc \hat W_2}{(X_1\wc X_1)^{1/2}}\ ,& \quad 
    \sQ_{\mathsf{BD},8} &= \hat W \wc \hat W_2 \ .
\end{align}

Finally the homogeneity of the correlator with respect to the vectors constraints the form,
\begin{align}
\begin{aligned}
	&\langle \CD^{(m)}_s (P_\alpha, \hat W)\,  \CO_{\Delta_1, l_1}(X_1, Z_1) \, \hat \CO_{\hat\Delta_2, j_2, s_2}(\hat X_2, \hat Z_2, \hat W_2)\rangle  \\ 
	&\qquad = \frac{\sQ_{\mathsf{BD}, 1}^{j_2}}{(X_1\wc X_1)^{(\Delta_1- \hat\Delta_2)/2 }\, (X_1\bc \hat X_2)^{\hat\Delta_2}}\, \sum_{\{n_k\}}N_{\mathsf{BD}, n_2\cdots n_8}\, \prod_{k=2}^8\, \sQ_{\mathsf{BD}, k}^{n_k} \ ,
\end{aligned}
\end{align}
where the exponents $n_k~(k=2,\cdots, n_8)$ are non-negative and subject to the three constraints due to the homogeneity,
\begin{align}
	\begin{aligned}
         2n_2 + n_3 + n_4 + n_5 &= l_1 - j_2 \ ,\\
        n_5 + n_7 + n_8 &= s_2 \ ,\\
        n_4 + n_6 + n_8 &= s \ .
	\end{aligned}
\end{align}
It follows that the correlator vanishes when $l_1 < j_2$.

\subsection{Two-point function of bulk local operators}\label{sec:BB}
Next we want to evaluate the correlator of two bulk operators with a spinning defect,
\begin{align}
\langle \CD^{(m)}_s (\hat W)\,  \CO_{\Delta_1, l_1}(X_1, Z_1) \, \CO_{\Delta_2, l_2}(X_2, Z_2)\rangle \ .
\end{align}
In this case, the correlation function can be fixed at most up to 
a function $f(\xi_1, \xi_2)$ of two cross ratios built from only the position vectors $X_i~(i=1,2)$ that are invariant under the residual conformal symmetry,
\begin{align}
	\xi_1 \equiv \frac{X_1 \wc X_2}{\left[ ( X_1 \wc X_1)(X_2 \wc X_2)\right]^{1/2}} \ , \qquad \xi_2 \equiv \frac{X_1 \bc X_2}{\left[ ( X_1 \wc X_1)(X_2 \wc X_2)\right]^{1/2}} \ .
\end{align}

To begin with, we consider correlators for a scalar defect ($s=0$) \cite{Billo:2016cpy} whose linearly independent invariant basis is given by $X_i \wc X_i$ for $i=1,2$ and
\begin{align}
\begin{aligned}
\sQ_{\mathsf{BB},1} &= \frac{X_1 \bc C_1 \wc X_2}{(X_1\wc X_1)\, (X_2\wc X_2)^{1/2}}\ , \qquad &
\sQ_{\mathsf{BB},2} &= \frac{X_2 \bc C_1 \wc X_2}{(X_1\wc X_1)^{1/2}\, (X_2\wc X_2)}\ ,\\
\sQ_{\mathsf{BB},3} &= \frac{X_1 \bc C_2 \wc X_2}{(X_1\wc X_1)^{1/2} \,(X_2\wc X_2)}\ , \qquad &
\sQ_{\mathsf{BB},4} &= \frac{X_1 \bc C_2 \wc X_1}{(X_1\wc X_1)\, (X_2\wc X_2)^{1/2}}\ , \\
\sQ_{\mathsf{BB},5} &= \frac{X_1 \bc C_1 \wc C_2 \bc X_2}{(X_1\wc X_1)\, (X_2\wc X_2)}\ , \qquad & 
\sQ_{\mathsf{BB},6} &= \frac{X_2 \wc C_1\bc C_2 \wc X_2}{(X_1\wc X_1)^{1/2}\, (X_2\wc X_2)^{3/2}}\ , \\
\sQ_{\mathsf{BB},7} &= \frac{C_1^{AI}\,C_{1AI}}{(X_1\wc X_1)}\ , & \qquad 
\sQ_{\mathsf{BB},8} &= \frac{C_2^{AI}\,C_{2AI}}{(X_2\wc X_2)}\ . &
\end{aligned}
\end{align} 
In addition, there are a few more invariants including the auxiliary vector $\hat{W}$ for a spinning defect, whose basis can be spanned by
\begin{align}
\begin{aligned}
&\hat{W} \wc C_1 \bc C_2 \wc \hat{W}\ , &\qquad 
&\hat{W} \wc C_1 \bc C_2 \wc X_i\ , &\qquad & X_i \wc C_1 \bc C_2 \wc \hat{W}\ ,\\
& X_i \bc C_1 \wc \hat{W}\ , &\qquad
& X_i \bc C_2 \wc \hat{W}\ ,  &\qquad
& X_i \wc \hat{W}\ . 
\end{aligned}
\end{align}
These are not linearly independent due to the identities \eqref{DBB_Identities}, \eqref{appendix_4} and \eqref{DBB_Identities2}, 
and we choose the following set,\footnote{
In the version 1, we included the invariant $\hat{W} \wc C_1 \bc C_2 \wc \hat{W}$ as a linearly independent basis, but it can be decomposed by the others due to the identity \eqref{appendix_4}. 
We thank Sunny Guha and Balakrishnan Nagaraj for informing us of this point.
}
\begin{align}
\begin{aligned}
\sQ_{\mathsf{BB},9} & = \frac{X_1 \bc C_1 \wc \hat{W}}{(X_1\wc X_1)} \ , & \qquad
\sQ_{\mathsf{BB},10} & = \frac{X_1 \bc C_2 \wc \hat{W}}{(X_1\wc X_1)^{1/2}\, (X_2\wc X_2)^{1/2}} \ , \\
\sQ_{\mathsf{BB},11} & = \frac{X_1 \wc \hat{W}}{(X_1\wc X_1)^{1/2}} \ , & \qquad 
\sQ_{\mathsf{BB},12} & = \frac{X_2 \wc \hat{W}}{(X_2\wc X_2)^{1/2}} \ .
\end{aligned}
\end{align}

In this linearly independent basis, the bulk two-point function takes the form,
\begin{align}
\langle \CD^{(m)}_s (\hat W)\,  \CO_{\Delta_1, l_1}(X_1, Z_1) \, \CO_{\Delta_2, l_2}(X_2, Z_2)\rangle = \sum_{\{n_k\}}\, f_{n_1\cdots n_{12}}(\xi_1, \xi_2)\,\frac{\prod_{k=1}^{12}\sQ_{\mathsf{BB},k}^{n_k}}{(X_1\wc X_1)^{\Delta_1/2}(X_2\wc X_2)^{\Delta_2/2}}\ .
\end{align}
The homogeneity conditions with respect to the spins of a spinning defect and bulk local operators give rise to three constraints on
the non-negative exponents $n_k~(k=1,\cdots, 12)$,
\begin{align}
	n_{9} + n_{10} + n_{11} + n_{12} &= s \ ,\\
    n_1 + n_2 + n_5 + n_6 + 2n_7 + n_{9}	&= l_1 \ ,\\
    n_3 + n_4 + n_5 + n_6 + 2n_8  + n_{10}    &= l_2 \ ,
\end{align}
leaving nine parameters in the correlator.

\subsection{Conserved current}
A spin-$l$ conserved current $J_l$ has to have dimension $\Delta = d - 2 + l$ to satisfy the conservation law in the index-free notation \cite{Costa:2011mg},
\begin{align}
	\partial^M\,D_M\, J_l (X, Z) = 0 \ ,
\end{align}
where $\partial_M \equiv \partial/\partial X^M$ and $D_M$ is the Todorov operator.
The conservation law imposes a set of constraints between the coefficients $N_n$ in the one-point function \eqref{bulk_one_point},
\begin{align}\label{conserved_constraint}
	a_{n-1}\,N_{n-1} + b_n\, N_n + c_{n+1}\,N_{n+1} &= 0 \ ,\qquad  \text{max}\left( 0, \left\lceil\frac{l-s}{2}\right\rceil\right) \le n \le \left\lfloor \frac{l}{2}\right\rfloor  \ ,
\end{align}
where $N_n$ with $n$ out of the range are to be understood as zero and
\begin{align}
	\begin{aligned}
		a_{n} &= - \frac{(l-2n)(l-2n-1)(l-2n-2)}{4} \ , \\
        b_n &= \frac{l-2n}{2}\left[ l^2 + (2m - 4n + s - 7)l + 8n^2 + 2(d+s+2 - m)n + d(m+s-2) -4m +10\right] \ ,\\
        c_{n} &= -2n (l-2n -s)(d+2n - m -1) \ .
	\end{aligned}
\end{align}
For example, the one-point function of a spin-$1$ conserved current $J_1$ vanishes due to the conservation law.

The correlation functions of a stress tensor $T \equiv J_2$ with $\Delta = d$ are determined uniquely for $s= 0$ and $s=1$ and satisfy the conservation law automatically, 
\begin{align}
		\langle \CD^{(m)}_{s=0}\, T(X, Z)\rangle = \frac{\sQ_3}{(X\wc X)^{d/2}} \ , \qquad
    	\langle \CD^{(m)}_{s=1}(\hat W)\, T(X, Z)\rangle = \frac{\sQ_1\,\sQ_3}{(X\wc X)^{d/2}} \ ,
\end{align}
but they are not unique for spin $s\ge 2$ with two unknown parameters $N_0$ and $N_1$.
Solving the constraint equations from the conservation law \eqref{conserved_constraint} we can fix the correlator up to an overall factor,
\begin{align}\label{One_point_StressTensor}
	\langle \CD^{(m)}_{s}(\hat W)\, T(X, Z)\rangle = \frac{1}{(X\wc X)^{d/2}} \left[ \sQ_1^{s-2}\,\sQ_2^2 + \frac{s- d(m+ s-2)}{2s\, (d+1-m)}\, \sQ_1^{s}\,\sQ_3 \right] \ .
\end{align}

\subsection{Recursion relations for the bulk one-point functions}\label{ss:Recursion_one-point}

Correlators of spinning local operators are known to follow from correlators of scalar operators acted on recursively by differential operators \cite{Costa:2011dw,Karateev:2017jgd}.
Similarly we expect the one-point functions of higher defect spins can be obtained by acting a certain type of differential operators on the one-point functions of lower defect spins.
Such differential operators should be symmetric scalar polynomials of the auxiliary vector $\hat W$ for a spinning defect so as to increase the defect spin, and contain the frame vectors $P_\alpha$ and their derivatives with the frame indices $\alpha$ contracted to respect the $\SO(m)$ invariance.
A moment's thought shows there are no $\SO(m)$ invariant differential operators linear in $\hat W$, and we must look for operators containing at least two $\hat W$'s.
Reminding that $\hat W$ is a transverse null vector, $\hat W \wc \hat W=0$, we find one first-order differential operator,
\begin{align}\label{1st_diff}
	\Fd_1 \equiv (P_\alpha\cdot \hat W)(P_\beta\cdot \hat W)\, \left(P_\alpha \cdot \frac{\partial}{\partial P_\beta}\right) \ ,
\end{align}
and two linearly independent second-order differential operators,
\begin{align}\label{2nd_diff}
	\begin{aligned}
    \Fd_2 &\equiv (P_\alpha\cdot \hat W)(P_\beta\cdot \hat W)\, \left(P_\gamma \cdot \frac{\partial}{\partial P_\alpha}\right)\,\left(P_\gamma \cdot \frac{\partial}{\partial P_\beta}\right) \ ,\\
    \Fd_3 &\equiv (P_\alpha\cdot \hat W)(P_\beta\cdot \hat W)\, \left(P_\alpha \cdot \frac{\partial}{\partial P_\gamma}\right)\,\left(P_\beta \cdot \frac{\partial}{\partial P_\gamma}\right) \ ,
    \end{aligned}
\end{align}
all of which are symmetric under the exchange of the two $\hat W$'s and raise the defect spin by two.

We will apply these differential operators to \eqref{bulk_one_point} to derive recursion relations between the one-point functions of different defect spins.
To simplify the notation, let us denote the invariants appearing in the one-point function \eqref{bulk_one_point} by
\begin{align}
	 [s,l,n;\Delta] \equiv \frac{ \sQ_1^{s-l + 2n}\, \sQ_2^{l-2n}\, \sQ_3^{n}}{(X \wc X)^{\Delta /2}} \ .
\end{align}
These invariants span the linearly independent basis for fixed $s$ and $l$, and are parameterized by $n$ with the range \eqref{Range_of_n} depending on the defect spin.

In this notation, the first-order differential operator $\Fd_1$ acts on the basis as
\begin{align}\label{D1action}
 \Fd_1\, [s,l,n;\Delta] = -(\Delta + s+ l - 2n)\,[s+2,l,n;\Delta] - 4n\, [s+2,l,n-1;\Delta]\ .
\end{align}
Here the invariants should be understood to be zero in the right hand side if their arguments are outside the range \eqref{Range_of_n}.
Similarly the actions of the second-order differential operators are 
\begin{align}\label{D3action}
	\begin{aligned}
	\Fd_2\,[s,l,n;\Delta] &= -\frac{(l-2 n) (l-2n-1)}{2}\,[s+2,l,n+1;\Delta]\\
    	&\qquad + \left(8 n^2 + 2n(2s-2l-1)-2s l + (\Delta+s)(\Delta +1-s) \right)\,[s+2, l, n;\Delta]\\
    	&\qquad -8n (n+s-1)\,[s+2, l, n-1;\Delta] \ ,
    \end{aligned}
\end{align}
and
\begin{align}
	\begin{aligned}
	\Fd_3\,[s,l,n;\Delta] &= -\frac{(l-2 n) (l-2n-1)}{2}\,[s+2,l,n+1;\Delta]\\
    	&\qquad + \left(8 n^2 +2 n (m+2 s-2l-1) -l (m+2
   s)+(\Delta +s) (\Delta
   +1-m-s)\right)\,[s+2, l, n;\Delta]\\
    	&\qquad -4n (m+2 (n+s-1))\,[s+2, l, n-1;\Delta] \ .
    \end{aligned}
\end{align}
The right hand side of \eqref{D3action} is independent of the codimension ($m$) of the defect in contrary to the action of $\Fd_3$, so we will use $\Fd_1$ and $\Fd_2$ to construct the spin-$(s+2)$ invariant basis from  the spin-$s$ basis through the recursion relations \eqref{D1action} and \eqref{D3action}.
One can also check they commute with each other,
\begin{align}
	[\Fd_1, \Fd_2] = 0 \ ,
\end{align}
thus we do not have to care about the ordering of the differential operators in constructing the higher spin basis.

We shall prove that the linearly independent basis $[s+2,l,n;\Delta]$ for the defect spin-$(s+2)$ correlator can be derived from $[s,l,n;\Delta]$ through the recursion relations.
Actually it is obvious such a construction is possible for $s\ge l-1$ as there are the same number of the linearly independent invariant basis for spin $s$ and $s+2$.
Less obvious is the case for $s\le l-2$ where the range \eqref{Range_of_n} of the parameter $n_s$ depends on $s$ as,
\begin{align}
	n_{s,\text{min}}\le n_s \le n_\text{max} \ ,\qquad n_{s,\text{min}}\equiv \left\lceil\frac{l-s}{2}\right\rceil \ ,\quad n_\text{max}\equiv \left\lfloor \frac{l}{2}\right\rfloor \ .
\end{align}
Hence the parameter $n$ ranges from $n_{s,\text{min}}-1\le n_{s+2} \le n_\text{max}$ for spin $s+2$,
and the number of the basis increases by one in going from spin $s$ to $s+2$.
We want to show the invariant basis $[s+2,l,n;\Delta]$ are uniquely determined from the basis $[s,l,n;\Delta]$ even in this case.
First, applying the two recursion relations \eqref{D1action} and \eqref{D3action} for the invariant with $n=n_\text{max}$,
one can represent $[s+2,l,n_\text{max};\Delta]$ and $[s+2,l,n_\text{max}-1;\Delta]$ by linear combinations of $\Fd_1\,[s,l,n_\text{max};\Delta]$ and $\Fd_2\,[s,l,n_\text{max};\Delta]$.
Then we can recursively use the relations to represent all the spin-$(s+2)$ invariants $[s+2,l,n;\Delta]$ for $n\le n_\text{max}-2$ by the spin-$s$ invariants $[s,l,n+1;\Delta]$ multiplied by the differential operators $\Fd_1$ and $\Fd_2$ appropriately.

We demonstrate the procedure in the simplest case with $s=0$ and $l=2k$ for $k\ge 1$, where the parameter $n_s$ ranges for $n_0 = k$ and $k-1\le n_2 \le k$.
There is only one invariant $[0,2k,k; \Delta ]$ for $s=0$ and two invariants $[2,2k,k-1; \Delta ], [2,2k,k; \Delta ]$ for $s=2$, which are related by
the recursion relations,
\begin{align}
	\begin{aligned}
		\Fd_1\, [0,2k,k; \Delta ] &= - \Delta\,[2,2k,k; \Delta ] - 4k\,[2,2k,k-1; \Delta ] \ ,\\
        \Fd_2\, [0,2k,k; \Delta ] &= \left( -2k + \Delta (\Delta +1)\right)\,[2,2k,k; \Delta ] - 8k (k-1)\, [2,2k,k-1; \Delta ] \ .
	\end{aligned}
\end{align}
Inverting the relations we find the differential basis for $s=2$,
\begin{align}\label{Recursion_for_s=2}
	\begin{aligned}
		[2,2k,k; \Delta ] &= -\frac{1}{(\Delta -1)(\Delta + 2k)} \left[ 2(k-1)\,\Fd_1\, [0,2k,k; \Delta ] -\Fd_2\, [0,2k,k; \Delta ] \right] \ , \\
        [2,2k,k-1; \Delta ] &= \frac{1}{4k (\Delta -1) (\Delta +2)} \left[ (2k - \Delta (\Delta +1))\, \Fd_1\,[0,2k,k; \Delta ] - \Delta \,\Fd_2\, [0,2k,k; \Delta ] \right] \ .
	\end{aligned}
\end{align}

In summary, we have shown that the bulk one-point function of a spin-$l$ primary operator with a spin-$s$ defect can be built recursively by acting differential operators on the one-point function with a scalar defect for $s$ even and with a spin-one defect for $s$ odd,
\begin{align}\label{Recursion_One}
	\langle \CD_s^{(m)}(\hat W)\, \CO_{l,\Delta}(X,Z)\rangle  = 
    		\FD_{s-s_0} (\hat W)\, \langle \CD_{s_0}^{(m)}\, \CO_{l,\Delta}(X,Z)\rangle \ ,
\end{align}
where $s_0 = s\, \text{mod}\,2$ and $\FD_s (\hat W)$ is a differential operator increasing the defect spin by $s$.

\section{Spinning defect OPE blocks} \label{ss:SDOPE}

\begin{figure}[tp]
\centering
\begin{tikzpicture}
	\shade[ball color=orange!50!red!50, opacity=0.70] (0,0) circle (1cm);
    \draw[dashed] (0,0) circle (2cm);
    \node[fill=white] at (1.2, 1.3) {$\CD^{(m)}_s$};
    
    \node at (3,0) {\Large $=$};
    
    \node at (5,0) {\Large $\sum_n c_{\CO_n}^{(m)} \,R^{\Delta_n}$};
    \node at (6.7,-0.5) {\Large $\CO_n$};
    \fill (6.7,0) circle (2pt);
    \node at (9,0) { $+ \quad \textrm{(descendants)}$};
\end{tikzpicture}
\caption{A spherical conformal defect of radius $R$ has the OPE in terms of bulk local operators located at the center.}
\label{fig:Defect-Exp}
\end{figure}
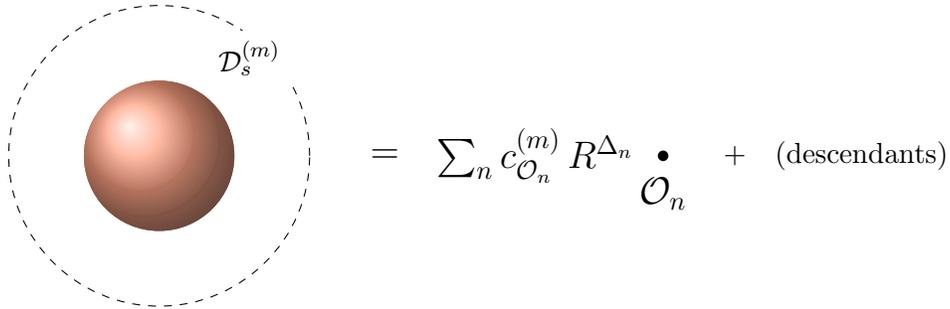

The OPE of defects by bulk local operators has beneficial applications to studying correlation functions of loop and surface operators \cite{Berenstein:1998ij,Corrado:1999pi,Gomis:2009xg,Chen:2007zzr}.
The structure of the OPE is fixed by the conformal symmetry, and amounts to the decomposition of defects into the conformal multiplets labeled by primary operators (see figure \ref{fig:Defect-Exp} for an illustration).
For example, a scalar conformal defect $\CD^{(m)}$ 
is expanded in the form \cite{Gadde:2016fbj,Fukuda:2017cup},
\begin{align}
	\CD^{(m)} = \sum_n\, \CB^{(m)} [\CO_n] \ ,
\end{align}
where $\CB^{(m)} [\CO_n]$ is the defect OPE (DOPE) block for a primary operator $\CO_n$ labeled by an irreducible representation $n$ of the conformal group.
The DOPE block $\CB^{(m)} [\CO_n]$ contains all contributions from the conformal multiplet of the primary operator, hence non-local function of $\CO_n$.

Now we would like to introduce the OPE of a spinning conformal defect in a similar manner to the scalar case \cite{Fukuda:2017cup}, and decompose it into the spinning DOPE blocks,
\begin{align}\label{Spinning_DOPE}
	\CD^{(m)}_s (\hat W) = \sum_n\, \CB^{(m)}_s [\CO_n, \hat W] \ .
\end{align}
One way to derive this form is to use the spectral decomposition of the identity operator,
\begin{align}\label{Spectral_Decomposition}
	{\bf 1} = \sum_n \, |\CO_n| \ ,
\end{align}
where $|\CO_n|$ is a projector onto the conformal multiplet of a primary operator $\CO_n$.
One can represent a correlation function with a spinning defect using either the blocks \eqref{Spinning_Defect} or the spectral decomposition of the identity operator \eqref{Spectral_Decomposition}.
By comparing the two forms, we find
\begin{align}\label{Block_Spectral}
	\langle \CB^{(m)}_s [\CO_n, \hat W] \cdots \rangle = \langle \CD^{(m)}_s (\hat W)\, |\CO_n| \cdots \rangle \ .
\end{align}
In the shadow formalism, $|\CO_n|$ is represented as a conformal integral of a primary $\CO_n$ of dimension $\Delta$ and its shadow operator $\tilde \CO_n$ with dimension $\tilde\Delta = d- \Delta$ \cite{Ferrara:1973eg,Ferrara:1972xe,Ferrara:1973vz,Ferrara:1972uq,Ferrara:1972ay,SimmonsDuffin:2012uy} (see appendix \ref{app:shadow} for more details).
For instance, the projector onto the conformal multiplet of a spin-$l$ primary operator $\CO_{\Delta, l}$ is given by
\begin{align}
	|\CO_{\Delta, l}| \equiv \frac{1}{\CN_{\Delta, l}}\,\int D^d X\, |\tilde \CO_{d-\Delta, l} (X, D_Z)\rangle \, \langle\CO_{\Delta, l}(X, Z) |\ .
\end{align}
Plugging into the equality \eqref{Block_Spectral}, 
we can read off the integral representation of the spinning DOPE block for a spin-$l$ operator,
\begin{align}\label{SDOPE_block}
	\CB^{(m)}_s [\CO_{\Delta, l}, \hat W] = \frac{1}{\CN_{\Delta, l}}\,\int D^d X\, \tilde \CO_{d-\Delta, l} (X, D_Z) \, \langle \CD^{(m)}_s (\hat W)\, \CO_{\Delta, l}(X, Z) \rangle \ .
\end{align}
This integral form allows us to deduce a few important aspects of spinning DOPE blocks from the properties of the bulk one-point function as we will show below.\footnote{To ensure that DOPE block has an appropriate behavior in the small radius limit, we have to impose a monodromy condition under the  frame vectors \cite{SimmonsDuffin:2012uy,Fukuda:2017cup}.}

\subsection{Casimir equation}
The bulk one-point function \eqref{bulk_one_point} is invariant under the entire conformal group $\SO (d+1, 1)$ generators $\hat J_{MN}$ that consist of two generators, $\hat J_{MN}^{(\CD)}, \hat J_{MN}^{(\CO)}$, acting on the defect and the bulk primary operator, respectively,
\begin{align}
	\left(\hat J_{MN}^{(\CD)} + \hat J_{MN}^{(\CO)} \right)\, \langle \CD^{(m)}_s(\hat W)\, \CO_{\Delta,l}(X,Z) \rangle = 0 \ .
\end{align}
Hence acting the quadratic Casimir operator for a spinning defect on the one-point function amounts to the quadratic Casimir equation for a primary operator, 
\begin{align}\label{Casimir_eq_DO}
	\hat J^{(\CD)\,2}\, \langle \CD^{(m)}_s(\hat W)\, \CO_{\Delta,l}(X,Z) \rangle = -
    \,\CC_{\Delta, l}\, \langle \CD^{(m)}_s(\hat W)\, \CO_{\Delta,l}(X,Z) \rangle \ ,
\end{align}
where $\CC_{\Delta, l} = \Delta (\Delta -d) + l (l+d-2)$ is the eigenvalue of the Casimir operator for a spin-$l$ operator.
Similarly the bulk one-point function for any primary operator $\CO_n$ satisfies the quadratic Casimir equation of the form \eqref{Casimir_eq_DO} with the Casimir eigenvalue $\CC_n$.

Now it is easy to translate the relation \eqref{Casimir_eq_DO} into the quadratic Casimir equation for the spinning DOPE block in the integral representation \eqref{SDOPE_block} as the generator $\hat J^{(\CD)}_{MN}$ does not act on the integrand except the one-point function.
To highlight the fact that $\hat J^{(\CD)}_{MN}$ only acts on a defect, we split the conformal generators into the orbital part $L_{MN}(P_\alpha)$ and spin part $S_{MN}(\hat W)$,
\begin{align}
	\hat J_{MN}^{(\CD)}\equiv L_{MN}(P_\alpha) + S_{MN}(\hat W) \ ,
\end{align}
which are realized as differential operators in the embedding space as \eqref{Orbital_generator} and ,
\begin{align}
	S_{MN}(\hat W) = \hat W_M\, \frac{\partial}{\partial \hat W^N} - \hat W_N\, \frac{\partial}{\partial \hat W^M} \ .
\end{align}
It is clear that these operators act only on the defect variables $P_\alpha,~\hat W$, hence the quadratic Casimir equation for the spinning DOPE block becomes,
\begin{align}\label{Casimir_Block}
	(L^2  + L^{MN}S_{MN} + S^2)\, \CB^{(m)}_s [\CO_n, \hat W] = - \CC_n\,\CB^{(m)}_s [\CO_n, \hat W] \ .
\end{align}

This equation appears to be independent of the spin of the block at first, but the spin dependence becomes manifest if we substitute the eigenvalue $-s(s+m -2)$ of 
the spinning part of the Casimir operator $S^2 = S^{MN}S_{MN}/2$.
Similarly the cross term operator $L^{MN}S_{MN}$ can be written as
\begin{align}
\begin{aligned}
	L^{MN}S_{MN} &= 2 (P_\alpha\cdot \hat W)\,(\partial_{P_\alpha}\cdot\partial_{\hat W}) - 2 P_\alpha^M \hat W^N\,\partial_{P_\alpha^N}\partial_{\hat W^M}\ ,\\
     	&= 2 \left[ (P_{\alpha}\cdot \hat W)\,(\partial_{P_{ \alpha}}\cdot\partial_{\hat W}) - (\hat W\cdot \partial_{P_{\alpha}}) ( P_{\alpha}\cdot \partial_{\hat W})\right] + 2m \,(\hat W\cdot \partial_{\hat W})\ ,\\
        &= 2 (P_{\alpha}\cdot \hat W)\,(\partial_{P_{ \alpha}}\cdot\partial_{\hat W}) + 2m s \ ,
\end{aligned}
\end{align}
where we used the shift symmetry $\hat W \to \hat W + \lambda\, P_\alpha$, implying
\begin{align}
	 P_{\alpha}\cdot \partial_{\hat W}\,\CD_s^{(m)}(\hat W) = 0 \ ,
\end{align}
in the correlator.
Hence the quadratic Casimir equation for the spinning block \eqref{Casimir_Block} becomes
\begin{align}\label{Casimir_DOPEB}
	\left[L^2  + 2 (P_{\alpha}\cdot \hat W)\,(\partial_{P_{ \alpha}}\cdot\partial_{\hat W}) \right]\, \CB^{(m)}_s [\CO_n, \hat W] = - \left[\CC_n + s(s-m -2)\right]\,\CB^{(m)}_s [\CO_n, \hat W] \ .
\end{align}

The Casimir equation for the conformal blocks of four local operators is shown to be equivalent to the Schr{\"o}dinger equation of the Calogero-Sutherland model
\cite{Isachenkov:2016gim,Isachenkov:2017qgn,Schomerus:2016epl,Schomerus:2017eny}.
It is intriguing to see if the Casimir equation for the spinning DOPE block \eqref{Casimir_DOPEB} has an analogous interpretation as the Schr{\"o}dinger equation of an integrable system.

\subsection{Recursion relation}
Combined with the integral representation \eqref{SDOPE_block}, the recursion relation for the one-point function presented in \eqref{Recursion_One} allows to construct the spinning DOPE block of a spin-$l$ operator from a block of lower spin,
\begin{align}\label{Recursion_DOPEB}
	\CB^{(m)}_s [\CO_{\Delta, l}, \hat W] = \FD_{s-s_0} (\hat W)\, \CB^{(m)}_{s_0} [\CO_{\Delta, l}] \ .
\end{align}

Acting the quadratic Casimir operator on the both hand sides of the recursion relation \eqref{Recursion_DOPEB}, 
one can show the action of the differential operator $\FD_{s-s_0}(\hat W)$  commutes with the action of the Casimir operator at least on the block of a spin-$l$ operator,
\begin{align}
	\begin{aligned}
	\hat J^{(\CD)\,2}\, \FD_{s-s_0} (\hat W)\, \CB_{s_0}^{(m)} [\CO_{\Delta, l}, \hat W] 
    &= - \CC_{\Delta, l}\,\FD_{s-s_0} (\hat W)\, \CB_{s_0}^{(m)} [\CO_{\Delta, l}, \hat W]\ ,\\
    	&= \FD_{s-s_0} (\hat W)\, \hat J^{(\CD)\,2}\,\CB_{s_0}^{(m)} [\CO_{\Delta, l}, \hat W] \ .
    \end{aligned}
\end{align}
The commutativity of the Casimir operator and $\FD_{s-s_0} (\hat W)$ would have been non-trivial to prove without the integral representation of the blocks, especially when they are represented as differential operators.

From the block decomposition of a spinning defect \eqref{Spinning_DOPE} and
the recursion relation \eqref{Recursion_DOPEB}, we would speculate an operator identity
\begin{align}\label{Defect_Recursion}
	\CD^{(m)}_s(\hat W) = \FD_{s-s_0} (\hat W)\, \CD_{s_0}^{(m)} \ .
\end{align}
If this is true, any correlation function in the presence of a spinning conformal defect can be built by acting a differential operator on a scalar or a spin one defect correlator.
It is interesting to examine if the identity \eqref{Defect_Recursion} holds in general.

\section{Correlation functions of two spinning defects} \label{ss:2-pt_of_defects}
In this section, we are concerned with the correlator of two spinning conformal defects,
\begin{align}
	\langle \CD^{(m_1)}_{s_1} (P_\alpha, \hat W_1)\, \CD^{(m_2)}_{s_2} (Q_\mu,\hat W_2)\rangle \ ,
\end{align}
where we make it explicit the dependence of the defects on the frame vectors.
We will enumerate the conformal invariants that the correlator can only depend on, then we decompose it into the channels labeled by the bulk primary operators using the spinning DOPE blocks.
We leverage the recursion relations \eqref{Recursion_DOPEB} to calculate the spin-$l$ channel of the spinning defect correlator from differential operators acting on a scalar defect correlator.
We demonstrate the procedure with a few simple examples.

\subsection{Cross ratios and invariant basis}
The correlator can be expanded by scalar invariants respecting the $\SO(m_1)$ and $\SO(m_2)$ symmetries rotating the frame vectors $P_\alpha~(\alpha = 1, \cdots m_1)$ and $Q_\mu~(\mu=1,\cdots, m_2)$, respectively.
Such invariants should be constructed from the projection matrices onto the normal planes to the respective defects,
\begin{align}
	\BP^{AB}\equiv P^{\alpha\, A}P_\alpha^B \ , \qquad \BQ^{AB} \equiv Q^{\mu\, A} Q_\mu^B \ .
\end{align}
The cross ratios are obtained by concatenating $\BP\cdot \BQ$ and taking the traces \cite{Gadde:2016fbj},
\begin{align}
	\eta_a &\equiv \text{Tr}\left[ (\BP\cdot \BQ)^a\right] \ .
\end{align}
The number of the cross ratios is equal to the rank of the matrix $\BP\cdot \BQ$ due to the Cayley-Hamilton theorem,
\begin{align}
	a= 1, \cdots, \text{min} (m_1, m_2, d+2-m_1, d+2-m_2) \ .
\end{align}

There are also conformal invariants including the auxiliary vectors $\hat W_i\, (i=1,2)$,
\begin{align}
\begin{aligned}
	\xi_a &\equiv \hat W_1\cdot \BQ\cdot (\BP\cdot \BQ)^a\cdot \hat W_1 \ , \\
    \chi_a &\equiv \hat W_2\cdot (\BP\cdot \BQ)^a\cdot \hat W_1 \ , \\
    \lambda_a &\equiv \hat W_2\cdot  (\BP\cdot \BQ)^a\cdot \BP\cdot\hat W_2 \ .
\end{aligned}
\end{align}
The cross ratios and the invariants are linearly independent basis for correlation functions of two conformal defects that are closed under the actions of the differential operators \eqref{1st_diff} and \eqref{2nd_diff}.
For example, $\Fd_1$ acts on the basis as
\begin{align}\label{d1actions}
\begin{aligned}
	\Fd_1(P_\alpha, \hat W_1)\,\eta_a &= 2\,a\, \xi_{a-1} \ , 
    \\
    \Fd_1(Q_\mu, \hat W_2)\,\xi_a &= 
    2\sum_{i=0}^{a}\,\chi_i\,\chi_{a-i} \ ,\\
    \Fd_1(P_\mu, \hat W_1)\,\xi_a &= 2\sum_{i=0}^{a-1}\,\xi_i\,\xi_{a-1-i} \ ,\\
    \Fd_1(P_\alpha, \hat W_1)\, \chi_a &= 2\sum_{i=0}^{a-1}\,\chi_i\,\xi_{a-1-i}\ ,
\end{aligned}
\end{align}
and $\Fd_2$ acts as
\begin{align}\label{d2actions}
\begin{aligned}
	\Fd_2(P_\alpha, \hat W_1)\,\eta_a &= 2a \left[ a\,\xi_{a-1} + \sum_{i=0}^{a-2}\,\xi_i\, \eta_{a-1-i}\right] \ ,\\
    \Fd_2(P_\alpha, \hat W_1)\,\chi_a &= (a+1)\chi_0\,\xi_{a-1} + 2a \sum_{i=1}^{a-1}\,\xi_i\, \eta_{a-1-i} + \sum_{i=0,j=1, i+j\le a-1}\,\chi_i\,\xi_{j-1}\,\eta_{a-i-j}\ ,\\
    \Fd_2(P_\alpha, \hat W_1)\,\xi_a &= 2\sum_{i=0}^{a-1}\left[ (2a + 1 - i)\,\xi_i\,\xi_{a-1-i} + \sum_{j=0}^{i-2} \xi_j\,\xi_{i-1-j}\,\eta_{a-j} \right] \ ,
\end{aligned}
\end{align}
and similarly for the other invariants.

Since the two-defect correlator is a function of the cross ratios and the conformal invariants,
it is straightforward to write the Casimir equation as a second-order partial differential equation with respect to these variables, while
we do not bother to write it explicitly to avoid the clutter.
We refer readers to \cite{Gadde:2016fbj} for the expression without defect spins.

\subsection{Integral representation by spinning DOPE blocks}
The correlator is seen to be decomposed into the channels of conformal multiplets through the DOPE block representation,
\begin{align}
	\begin{aligned}
	\langle \CD^{(m_1)}_{s_1} (\hat W_1)\, \CD^{(m_2)}_{s_2} (\hat W_2)\rangle 
    	&= \sum_n\, \langle \CB^{(m_1)}_{s_1} [\CO_n, \hat W_1]\,\CB^{(m_2)}_{s_2} [\CO_n, \hat W_2]\rangle \ .
	\end{aligned}
\end{align}
The contribution from a spin-$l$ primary operator of dimension $\Delta$ is read off by employing the integral representation \eqref{SDOPE_block} of the block, or directly inserting the projector between the two defects,
\begin{align}\label{IntegralRep_TwoDefects}
	\begin{aligned}
		\langle \CD^{(m_1)}_{s_1} (\hat W_1)\, \CD^{(m_2)}_{s_2} (\hat W_2)\rangle|_{\text{spin-}l}  &\\
        = &\int D^d X_1\, D^d X_2\, \langle \tilde\CO_{d-\Delta, l}(X_1, D_{Z_1})\,\tilde\CO_{d-\Delta, l}(X_2, D_{Z_2})\rangle\,\\
        	&\qquad \cdot \langle \CD^{(m_1)}_{s_1} (\hat W_1)\, \CO_{\Delta, l} (X_1, Z_1)\rangle\, \langle \CD^{(m_2)}_{s_2} (\hat W_2)\, \CO_{\Delta, l} (X_2, Z_2)\rangle \ ,\\
        = & \int D^d X\,\langle \CD^{(m_1)}_{s_1} (\hat W_1) \, \CO_{\Delta, l} (X, D_Z)\rangle\,
       \langle \CD^{(m_2)}_{s_2} (\hat W_2)\,\tilde\CO_{d-\Delta, l} (X, Z)\rangle \ .
	\end{aligned}
\end{align}
Substituting the bulk one-point function \eqref{bulk_one_point} into the integrand, we are left with performing the integral over $X$, which we find intractable even for simple cases.
Nonetheless, we believe this integral form is useful in studying the Mellin representation of defect correlators in the same spirit of the recent works \cite{Rastelli:2017ecj,Goncalves:2018fwx} on the Mellin amplitudes of correlation functions of local operators in defect CFTs.
We shall leave it for an interesting future work.

\subsection{Recursion relations}
We have seen so far the two-defect correlator may be obtained, in principle, by solving the quadratic Casimir equation \eqref{Casimir_DOPEB} in terms of the cross ratios and invariants, or performing the conformal integral of \eqref{IntegralRep_TwoDefects} derived in the shadow formalism.
Here we will introduce another recursive method to solve the spinning defect correlators by employing the recursion relation for the spinning DOPE block \eqref{Recursion_DOPEB}.

We assume for simplicity that the defect spins are \emph{even}, and apply the recursion relation to reduce the spinning defect correlator to a scalar defect correlator,
\begin{align}\label{Recursion_Defect_Correlator}
	\langle \CD^{(m_1)}_{s_1} (\hat W_1)\, \CD^{(m_2)}_{s_2} (\hat W_2)\rangle|_{\text{spin-}l} = \FD_{s_1}(\hat W_1)\,\FD_{s_2}(\hat W_2)\, \langle \CD^{(m_1)}\, \CD^{(m_2)}\rangle|_{\text{spin-}l} \ .
\end{align}
Scalar defect correlators are known in a closed form in certain cases by solving a hypergeometric differential equation \cite{Gadde:2016fbj}.
We will use \eqref{Recursion_Defect_Correlator} to fix spinning defect correlators in such cases below.

\subsubsection{Correlator with codimension-one defect}
We consider, as an illustrating example, a correlator of spinning defects of codimension $m_1=m$ and $m_2=1$.
The latter is a scalar defect as there are no spins for the transverse $\SO(1)$ group.
As is indicated by \eqref{bulk_one_point_codim_one}, only a scalar primary channel contributes to the correlator.
Thus we have the relation
\begin{align}
	\langle \CD^{(m)}_{s} (\hat W)\, \CD^{(1)} \rangle  = \FD_{s}(\hat W)\, \langle \CD^{(m)}\, \CD^{(1)}\rangle|_{\text{scalar}} \ .
\end{align}
The invariant basis for the scalar channel is spanned by $[s,0,0;\Delta]$ for any $s$, and the recursion relation \eqref{D1action} reduces the invariant with even $s$ to the one with $s=0$,
\begin{align}
	[s,0,0;\Delta] = (-1)^{s/2}\prod_{i=1}^{s/2} (\Delta + s - 2i)^{-1}\, \Fd_1^{s/2}\, [0,0,0;\Delta] \ .
\end{align}
Then we can read off the differential operator,
\begin{align}
	\FD_{s}(\hat W) = (-1)^{s/2}\prod_{i=1}^{s/2} (\Delta + s - 2i)^{-1} \, \Fd_1^{s/2}\ .
\end{align}
The scalar defect correlator is given by solving the quadratic Casimir equation \cite{Gadde:2016fbj},
\begin{align}
	\langle \CD^{(m)}\,\CD^{(1)}\rangle|_{\text{scalar}} = \eta_1^{-\Delta/2}\, {}_2 F_1 \left( \Delta/2, 1+ (\Delta -m)/2, 1 + \Delta -d/2; \eta_1^{-1}\right) \ ,
\end{align}
and is a function of the cross ratio.
By representing $\Fd_1$ as a differential operator acting on $\eta_1$ with a chain rule,
\begin{align}
	\Fd_1 = 2\, \xi_0\, \partial_{\eta_1} \ ,
\end{align}
we obtain the spinning defect correlator of even spin $s$ with a codimension-one defect,
\begin{align}
	\langle \CD^{(m)}_{s} (\hat W)\, \CD^{(1)} \rangle  = \left(\prod_{i=1}^{s/2} \frac{-2\xi_0}{\Delta + s - 2i} \right)\, \left(\frac{\partial}{\partial \eta_1}\right)^{s/2}\,\eta_1^{-\Delta/2}\, {}_2 F_1 \left( \Delta/2, 1+ (\Delta -m)/2, 1 + \Delta -d/2; \eta_1^{-1}\right)\ .
\end{align}

\subsubsection{Correlator of two codimension-two defects}
Our next example is the correlator of two codimension-two spinning defects.
When the two defects are scalar, the correlator is characterized by two cross ratios $\eta_1$ and $\eta_2$.
The spin-$l$ channel is fixed by solving the quadratic Casimir equation \cite{Gadde:2016fbj},
\begin{align}\label{Two_Scalar_Defects}
	\langle \CD^{(2)}(P_\alpha)\,\CD^{(2)}(Q_\mu)\rangle |_{\text{spin-}l} = (-1)^l\, \frac{x z}{x-z}\,\left[ k_{\Delta + l} (x)\,k_{\Delta -l -2}(z) - k_{\Delta + l} (z)\,k_{\Delta -l -2}(x)\right] \ ,
\end{align}
where $k_\beta (x)$ is given by the hypergeometric function
\begin{align}
	k_\beta (x) = x^{\beta/2}\, {}_2 F_1 \left(\beta/ 2, \beta/2, \beta; x \right) \ ,
\end{align}
and the new variables $x, z$ are related to the cross ratios by
\begin{align}\label{Cross_xz}
	\eta_1 = \frac{2(1+v)}{u}\bigg|_{u = xz, v=(1-x)(1-z)}\ , \qquad \eta_2 = \frac{2(1+6v + v^2)}{u^2}\bigg|_{u = xz, v=(1-x)(1-z)}\ .
\end{align}

We will focus on the case where one of the two defects has spin-$2$ and the other is a scalar defect,
\begin{align}
	\langle \CD^{(2)}_{2}(P_\alpha, \hat W)\, \CD^{(2)}(Q_\mu) \rangle \ .
\end{align}
One can determine the spin-$l$ channel of the defect correlator through the recursion relation \eqref{Recursion_Defect_Correlator} once the differential operator $\FD_2$ acting on the scalar defect correlator is given.
We can read off $\FD_2$ from \eqref{Recursion_for_s=2} relating the invariant basis of the one-point functions for $s=2$ and $s=0$.
For instance, the differential operator for the stress tensor channel follows from the recursion relation \eqref{Recursion_for_s=2} for the one-point functions \eqref{One_point_StressTensor} with $m=2$ and $\Delta = d$,
\begin{align}
	\langle \CD^{(2)}_2(\hat W)\,T(X,Z)\rangle \propto \left[ (d-1)\Fd_1 + \Fd_2\right]\,\langle \CD^{(2)}\,T(X,Z)\rangle \ ,
\end{align}
hence $\FD_2(\hat W) = (d-1)\Fd_1 + \Fd_2$.

In applying $\FD_2(\hat W)$ to the scalar defect correlator \eqref{Two_Scalar_Defects} with $l=2$, we use \eqref{d1actions} and \eqref{d2actions} to recast it into the form,
\begin{align}
	\FD_2(\hat W) = 2\left[ d\,\xi_0\,\partial_{\eta_1} + (2d\,\xi_1 + \xi_0\,\eta_1) \,\partial_{\eta_2}\right] \ .
\end{align}
Acting it on the scalar correlator \eqref{Two_Scalar_Defects} with the relations \eqref{Cross_xz} yields the stress tensor channel of the correlator of spin-$2$ and scalar defects,
\begin{align}
	\langle \CD^{(2)}_{2}(P_\alpha, \hat W)\, \CD^{(2)}(Q_\mu) \rangle|_{\text{spin-}2} = \FD_2(\hat W) \, \langle \CD^{(2)}(P_\alpha)\,\CD^{(2)}(Q_\mu)\rangle |_{\text{spin-}2} \ .
\end{align}

\section{Discussion}\label{ss:discussion}
We have undertaken the studies of spinning conformal defects with the hope of finding new structures of and constraints on correlation functions and the OPE in defect CFTs.
We found the residual symmetry preserved by a conformal defect is sufficient enough to determine the kinematic parts of correlation functions and allows us to derive the integral representation \eqref{SDOPE_block} of the spinning defect OPE block, which turns out to be useful to deduce the recursion relations for the correlator of two spinning defects from those for the bulk one-point functions.
In spite of the formal progresses we have made, our formulation lacks concrete examples of spinning conformal defects with which one can test the validity of our formulation.
One simplest construction of a spinning defect would be to smear a defect local operator with transverse spin over the worldvolume of a defect.

We should comment on a subtlety of describing a spinning defect in the dual frame with the dual frame vector $\tilde P_{\tilde\alpha}$.
In the dual frame, a defect is a function of $\tilde P_{\tilde\alpha}$ supported on the hypersurface spanned by the dual frame vectors, and the spin indices are contracted with the auxiliary vector $\hat W$ subject to the transversality condition,
\begin{align}
	\tilde P_{\tilde\alpha} \cdot \hat W = 0\ .
\end{align}
Then there are no differential operators corresponding to $\Fd_1, \Fd_2$ that raise the defect spins by two in the dual frame, while there is only one second-order differential operator we can construct from $\tilde P_{\tilde\alpha}$,
\begin{align}
	\tilde\Fd = \left(\hat W \cdot \frac{\partial}{\partial \tilde P_{\tilde \alpha}}\right)^2 \ .
\end{align}
We were able to construct a higher-spin basis from lower-spin ones by acting with the differential operators $\Fd_1$ and $\Fd_2$ in the original frame, but it is not clear if we can repeat the same construction in the dual frame.
One of the reasons for this difficulty would be that the conformal invariants appearing in the bulk one-point function \eqref{bulk_one_point} cannot be written without using the dot inner product in the index-split approach, which may intrinsically force us to stick to the original frame in describing spinning conformal defects.

There are a few future directions of interest a spinning conformal defect may come into play in.
In this paper, we are only concerned with conformal defects in the symmetric traceless tensor representations of the transverse rotation group $\SO (m)$, but it should be straightforward to extend our formulation to more general representations along the lines of the previous works on local operators \cite{Costa:2014rya,Costa:2016hju,Rejon-Barrera:2015bpa}.

Correlation functions in a scalar defect CFT have particularly nice analytic structures in the Mellin representation \cite{Goncalves:2018fwx,Rastelli:2017ecj},
and a surprising connection to integrable models
\cite{Isachenkov:2017qgn}.
It is not hard to speculate these properties are inherited to spinning defects correlators.

Finally we would like to note the spinning DOPE block could be viewed as a higher-spin field propagating on the moduli space $\CM^{(d,m)}$ that has a coset structure $\SO(d+1, 1)/ \SO(d+1-m, 1)\times \SO(m)$.
This would be a natural generalization of the case for the scalar DOPE block that also has the same moduli space as $\CM^{(d,m)}$.
This coset structure of $\CM^{(d,m)}$ allows a map from the scalar block into a scalar field on the AdS$_{d+1}$ space by the Radon transform \cite{Fukuda:2017cup} (see also \cite{Czech:2016xec,deBoer:2016pqk}).
Then it is tempting to expect that the spinning DOPE block is the Radon transform of a free higher-spin field on the AdS space. 
One however encounters an immediate obstruction to this identification as they have different index structures with respect to the spins.
One possibility to make it work is to use a projected higher-spin field onto a codimension-$m$ totally geodesic submanifold in AdS$_{d+1}$ as a Radon transform of the block for a spinning conformal defect of codimension-$m$ in CFT$_d$ (see \cite{Czech:2016tqr} for a related work).
We hope to address this issue with a view to understanding the holographic principle in the future.


\acknowledgments  
We would like to thank V.\,Gon\c{c}alves for the correspondence and valuable discussion, and K.\,Tamaoka for helpful comments.
The work of T.\,N. was supported in
part by JSPS Grant-in-Aid for Young Scientists (B) No.15K17628 and JSPS Grant-in-Aid
for Scientific Research (A) No.16H02182.
The work of N.\,K. was supported in part by the Program for Leading Graduate Schools, MEXT, Japan and also supported by World Premier International Research Center Initiative (WPI Initiative), MEXT, Japan.

\appendix
\section{Identities}
The product of two different gauge invariants takes the form,
\begin{align}
	(C_1 \bc C_2)^{IJ} = (X_1 \bc X_2)\, Z_1^I Z_2^J + (Z_1 \bc Z_2)\, X_1^I X_2^J - (X_1 \bc Z_2)\, Z_1^I X_2^J - (X_2\bc Z_1)\, X_1^I Z_2^J \ .
\end{align}
Then one can derive the following identities used for the bulk two-point functions in section \ref{sec:BB}:
\begin{align}\label{DBB_Identities}
\begin{aligned}
 (\hat{W} \wc C_1 \bc C_2 \wc \hat{W})(X_2 \wc X_i) - (\hat{W} \wc C_1 \bc C_2 \wc X_i )(X_2 \wc \hat{W}) &= (X_i \wc C_2 \wc \hat{W})(X_2 \bc C_1 \wc \hat{W})\ , \\
(\hat{W} \wc C_1 \bc C_2 \wc \hat{W})(X_1 \wc X_i) - (X_i \wc C_1 \bc C_2 \wc \hat{W})(X_1 \wc \hat{W}) &= (X_i \wc C_1 \wc \hat{W})(X_1 \bc C_2 \wc \hat{W})\ , 
\end{aligned}
\end{align}
for $i=1,2$.
We also find another useful identity, 
\begin{align}\label{appendix_4}
\begin{aligned}
	&(\hat W \wc C_1 \bc C_2 \wc \hat W)(X_2 \wc X_2)(X_1 \wc X_2)  \\
	 &\quad = (X_2 \wc C_1 \bc C_2 \wc X_2)(X_1 \wc \hat{W})(X_2 \wc \hat{W})   
	+ (X_1 \bc C_2 \wc \hat{W})(X_2 \wc C_1 \wc \hat{W})(X_2 \wc X_2) \\
	&\qquad + (X_2 \bc C_1 \wc \hat{W})(X_2 \wc C_2 \wc \hat{W} )(X_1 \wc X_2) + (X_2 \bc C_1 \wc X_2)(X_2 \wc C_2 \wc \hat{W})(X_1 \wc \hat{W})\ .
\end{aligned}
\end{align}
In addition, we find
\begin{align}\label{DBB_Identities2}
(X_1 \bc C_i \wc \hat{W})(X_i \bc X_2) - (X_2 \bc C_i \wc \hat{W})(X_1 \bc X_i) &= (X_i \wc \hat{W})(X_1 \bc C_i \bc X_2) \ .
\end{align}
Note that no summation is made in this identity.

\section{Shadow operator}\label{app:shadow}
A projector to the irreducible representation of the conformal group labeled by an operator $\CO_{\Delta, l}$ of dimension $\Delta$ and spin-$l$ is defined by \cite{SimmonsDuffin:2012uy}
\begin{align}
	|\CO_{\Delta, l}| \equiv \frac{1}{\CN_{\Delta,l}}\, \frac{1}{l! (d/2-1)_l } \, \int\,D^d X\, |\CO_{\Delta,l} (X, D_Z)\rangle\, \langle \tilde\CO_{\Delta,l} (X, Z)| \ ,
\end{align}
where $\tilde \CO$ is the shadow operator for $\CO$ of dimension $d-\Delta$ defined by
\begin{align}
	\tilde\CO_{\Delta,l} (X, Z) \equiv \frac{1}{\CN_{d-\Delta,l}}\, \frac{1}{l! (d/2-1)_l } \, \int\,D^d X\,\langle \CO_{\Delta,l} (X, Z)\,\CO_{\Delta,l} (Y, D_W)\rangle \big|_{\Delta \to d-\Delta}\, \CO_{\Delta,l} (Y, W) \ .
\end{align}
In order to fix the normalization constant, we normalize the two-point function in the following form,
\begin{align}
	\langle \CO_{\Delta,l} (X, Z)\, \CO_{\Delta,l} (Y, W) \rangle = \frac{\left[\text{Tr}\,(-C_{ZX}\cdot C_{YW})\right]^l}{(-2 X\cdot Y)^{\Delta + l}} \ .
\end{align}
Then we demand the shadow transform of the shadow operator brings back to the original operator, $\tilde{\tilde \CO} = \CO$, by tuning the constant to \cite{Dolan:2011dv,Sleight:2017fpc}
\begin{align}
	\CN_{\Delta, l} = \pi^{d/2}\,(d-\Delta -1)_l\, \frac{\Gamma(d/2 - \Delta)}{\Gamma (\Delta + l)} \ .
\end{align}
In this normalization, one can check the projector acts trivially in a correlator with $\CO$,
\begin{align}
	\begin{aligned}
		\langle \CO_{\Delta,l} (X, Z)\, |\CO_{\Delta,l}|\cdots\rangle &= \frac{1}{\CN_{\Delta,l}}\, \frac{1}{l! (d/2-1)_l } \, \int\,D^d Y\, \langle \CO_{\Delta,l} (X, Z)\,\CO_{\Delta,l} (Y, D_W)\rangle\, \langle \tilde\CO_{d-\Delta,l} (Y, W)\cdots\rangle \ , \\
        &= \frac{1}{\CN_{\Delta,l}}\, \frac{1}{l! (d/2-1)_l } \, \int\,D^d Y\, \langle \tilde\CO_{d-\Delta,l} (X, Z)\,\tilde\CO_{d-\Delta,l} (Y, D_W)\rangle\big|_{\Delta \to d-\Delta}\, \langle \tilde\CO_{d-\Delta,l} (Y, W)\cdots\rangle \ , \\
        &= \langle \CO_{\Delta,l} (X, Z)\cdots \rangle \ .
	\end{aligned}
\end{align}

\bibliographystyle{JHEP}
\bibliography{Spinning_Defect}

\end{document}